\documentstyle[epsf,amssymb]{article}

\renewcommand{\theequation}{\arabic{section}.\arabic{equation}}

\begin{document}
\begin{center}

%%%%%%%%%%%%%%%%%%%%%%%%%%%%%%%%%%%%%%%%%%%%%%%%

{\Large {\bf Quantum Phase Transition in Lattice Model of Unconventional 
Superconductors }}\\
\vspace{1cm}
{\Large Kenji Sawamura, Yuki Moribe, and
Ikuo Ichinose} \\
\vspace{1cm}
{Department of Applied Physics,
Nagoya Institute of Technology, Nagoya, 466-8555 Japan 
} 

{\bf abstract}   
\end{center}
In this paper we shall introduce a lattice model of unconventional
superconductors (SC) like $d$-wave SC in order to study 
quantum phase transition at vanishing temperature ($T$).
Finite-$T$ counterpart of the present model was proposed previously with which
SC phase transition at finite $T$ was investigated.
The present model is a noncompact U(1) lattice-gauge-Higgs model
in which the Higgs boson, the Cooper-pair field, is put on lattice links
in order to describe $d$-wave SC.
We first derive the model from a microscopic Hamiltonian in the path-integral
formalism and then study its phase structure by means of the Monte Carlo 
simulations.
We calculate the specific heat, monopole densities and the magnetic penetration
depth (the gauge-boson mass). 
We verified that the model exhibits a second-order phase transition from
normal to SC phases.
Behavior of the magnetic penetration depth is compared with that obtained
in the previous analytical calculation using XY model in four dimensions.
Besides the normal to SC phase transition, we also found that another
second-order phase transition takes place within the SC phase in the 
present model.
We discuss physical meaning of that phase transition.

\newpage

%%%%%%%%%%%%%%%%%%%%%%%%%%%%%%%%%%%%%%%%%%%%%%%%%%%%%%%%%%%%%%%%%
\section{Introduction}
Ginzburg-Landau (GL) theory plays a very important role in study
of the superconducting (SC) phase transition.
In particular GL models defined on a lattice have been extensively studied
in order to investigate effects of topological excitations like vortices
nonperturbatively.
These models are nothing but gauge-Higgs models on a lattice that
also play an important role in elementary particle physics.

At present many materials that have unconventional SC phase have
been discovered\cite{UCSC}.
Among them, some of the materials, including the high-temperature $(T)$
cuprates, the rare-earth heavy-fermion materials, etc, exhibit
$d$-wave SC.
Order parameter (the Cooper-pair wave function) of the $d$-wave SC
has nodes.
Therefore in order to formulate a GL theory of the $d$-wave SC,
the Cooper-pair field, i.e., the Higgs field, must be put on {\em links
of the lattice} instead of lattice sites. 

In the previous papers\cite{UV1,UV2}, we introduced a new type of 
lattice gauge-Higgs
models that describe $d$-wave SC phase transitions.
We call these models U-V models where the $U$-field refers to the ordinary
gauge field whereas the $V$-field is the Higgs field defined on links.
In Ref.\cite{UV1}, a compact-U(1)-gauge version was studied from the 
viewpoint of
the t-J model that is a canonical model of the high-$T_c$ cuprates.
In the slave-particle representation, the $U$-field is an emergent 
{\em compact}
U(1) gauge boson and the $V$-field is a spinon-pair field for the 
resonating-valence-bond (RVB) order.
By means of Monte Carlo (MC) simulations, we found that there appears a phase 
transition line as the coupling between the $U$ and $V$-fields is increased.
We measured instanton densities and expectation value of the Wilson loops,
and verified that the observed phase transition is the one from a 
confinement to Higgs phases.
There the confinement phase corresponds to
electron phase, whereas the Higgs phase to fractionalized spin-gap phase.
Numerical results show that the phase transition is of first order.

In Ref.\cite{UV2}, on the other hand, we studied finite-$T$ phase transition 
of unconventional SC by using the U-V model.
In this approach, the $U$-field is a {\em noncompact} U(1) gauge field 
corresponding to the usual electromagnetic field, whereas $V$ is the 
electron Cooper-pair field.
There we considered the London limit of $V$, i.e., $V_{x,i}\in $U(1).
Gauge-invariant gauge-boson mass (the inverse correlation length of the
magnetic field) is an order parameter for the normal to SC phase transition.
We observed the phase transition by means of the MC simulations and 
found that the transition is of first order.
However as the anisotropy of the layered structure is increased,
signal of the first-order phase transition is getting weaker, e.g.,
the discontinuity of the gauge-boson mass at the transition point 
is getting smaller.

In the present paper, we shall continue to study critical properties
of the unconventinal SC by using the U-V model.
In particular we focus on the quantum phase transition (QPT) at $T=0$.

In Section 2, we shall derive the GL model 
for the QPT starting from a microscopic Hamiltonian of the Cooper pair.
In Section 3, we shall show the numerical calculations of the U-V model
in $(2+1)$ dimensions ($(2+1)$D). 
We calculated the ``internal energy", ``specific heat", 
``monopole" density of the $U$ and $V$-fields, and the magnetic penetration 
depth (the inverse gauge-boson mass).
We found that the model exhibits a second-order phase transition from 
the normal to SC phases at certain critical value of the parameter $g_c$, 
which controls quantum fluctuations. (Larger $g$ suppresses quantum 
fluctuations more.)
In Section 4, we report the result of the MC simulations for the U-V model
in $(3+1)$D.
The reason why we study both the $(2+1)$ and $(3+1)$D models
is that most of the materials of the $d$-wave SC are quasi-2D
and have a layered structure.
We found that the $(3+1)$D model also exhibits a second-order SC phase 
transition.
However besides that transition, the model has another phase transition
at larger $g'_c>g_c$ within the SC phase.
This phase transition is also of second-order.
We shall discuss its physical meaning.
Section 5 is devoted for discussion.

%%%%%%%%%%%%%%%%%%%%%%%%%%%%%%%%%%%%%%%%%%%%%%%%%%
\setcounter{equation}{0}
\section{Ginzburg-Landau Model on a lattice}
In this section, we shall derive the lattice GL model from the 
microscopic Hamiltonian by using the path-integral formalism.
To this end, let us first introduce a Cooper-pair field $\Phi_{x,i}$ on 
links $(x,i)$ of $d$-dimensional hypercubic lattice with lattice
spacing $a_0(=1)$, where $x$ denotes
the lattice site and $i$ is the direction index $i=1,\cdots, d$.
In terms of the electron annihilation operator $\psi_{x,\sigma}$,
where $\sigma=\uparrow, \downarrow$ is the spin index,
\begin{equation}
\Phi_{x,i}= \psi_{x,\uparrow}\psi_{x+i,\downarrow}-
\psi_{x,\downarrow}\psi_{x+i,\uparrow}.
\label{C-P}
\end{equation}
In the $d$-wave SC's, amplitude of on-site Cooper pair is vanishingly small
because of, e.g., the strong on-site Coulomb repulsion.
Furthermore, for example, in the $d_{x^2-y^2}$-wave SC state, $\langle
\Phi_{x,1}\cdot \Phi_{x,2}\rangle<0$,
i.e., the Cooper-pair field changes its sign under a $\pi/2$ rotation 
in the $1-2$ plane of the real space.
One should also notice that SC phase transition in the underdoped region
of the high-$T_c$ materials can be understood as a Bose-Einstein condensation
of a bosonic bound state of a pair of charge carriers.
From these observations, we propose the following quantum Hamiltonian for
the $d$-wave SC's,
\begin{eqnarray}
H&=&-t\sum_{x,i\neq j}(\Phi_{x,i}U^\ast_{x+i,j}\Phi^\ast_{x+j,i}U^\ast_{x,j}
+ \mbox{H.c.}) 
-C\sum_{x,i\neq j}(\Phi_{x,i}\Phi^\ast_{x+i,j}\Phi_{x+j,i}
\Phi^\ast_{x,j}+ \mbox{H.c.}) \nonumber \\
&& +D\sum_{x,i\neq j}(\Phi_{x,i}\Phi^\ast_{x+i,j}U^\ast_{x+j,i}U^\ast_{x,j}
+ \mbox{H.c.}) 
+\sum_{x,y,i,j}{\cal V}_{x,y}(N_{x,i}-\rho_0)(N_{y,j}-\rho_0),
\label{H}
\end{eqnarray}
where $t,C,D$ are parameters,
the electromagnetic (EM) gauge field $U_{x,i}$ is related to 
the EM vector potential $\vec{A}^{\rm em}$ as $U_{x,i}=e^{iA_{x,i}},
\; A_{x,i}=\int^{x+i}_x\vec{A}^{\rm em}d\vec{\ell}$ and $N_{x,i}$
is the number operator of the Cooper-pair field, 
$N_{x,i}=\Phi^\ast_{x,i}\Phi_{x,i}$.
$\rho_0$ denotes an uniform background positive charge and ${\cal V}_{x,y}$ 
is a Coulomb potential including long-range interactions like,
\begin{equation}
{\cal V}_{x,y}=\left\{
    \begin{array}{ll}
    e^2V_0\; ,& x=y\\
    e^2/|x-y|,&  |x-y|\gg 1
    \end{array}
\right.
\label{V}
\end{equation}
for $d=3$.
As we are interested in the long-range interaction in Eq.(\ref{V}) in the later
discussion, we have ignored the dependence on direction of the Cooper pairs
in the Coulomb potential.
The $t$-term in Eq.(\ref{H}) is noting but the hopping of the Cooper pair
and the $C$-term is a plaquette term of the link field $\Phi_{x,i}$
that stabilizes the phase fluctuations of $\Phi_{x,i}$.
On the other hand, the $D$-term determines relative phase of adjacent
$\Phi$'s.
The Hamiltonian (\ref{H}) might be derived from a canonical model of the
high-$T_c$ cuprates like the t-J model
by integrating out electrons just like Gorikov's
derivation of the GL Hamiltonian for the conventional $s$-wave
SC \cite{Gorikov}\cite{IM}. 
It is obvious that the Hamiltonian $H$ in Eq.(\ref{H}) is invariant
under the following local gauge transformation,
\begin{equation}
U_{x,i} \rightarrow e^{i\omega_{x+i}}U_{x,i}e^{-i\omega_x}, \;\;\;
\Phi_{x,i} \rightarrow e^{-i\omega_{x+i}}\Phi_{x,i}e^{-i\omega_x}.
\label{localGT}
\end{equation}
In the rest of this section, we shall derive the GL model from the
Hamiltonian (\ref{H}).
If the reader is not interested in the derivation, please go to the 
result Eq.(\ref{action}) directly.

We employ the path-integral formalism by introducing imaginary time $\tau$.
Then the partition function $Z$ is given by
\begin{equation}
Z=\int [d\Phi dA]e^{-S}, 
\label{Z}
\end{equation}
with the following action $S$ 
\begin{equation}
S=\int d\tau \Big[\sum_{x,i}\Phi^\ast_{x,i}\partial_\tau \Phi_{x,i}
+H(\Phi,U)\Big]+S_{\rm em}(A),
\label{S}
\end{equation}
where 
$$
S_{\rm em}(A)\propto \int d\tau \sum F^2_{ij}(x)
=\int d\tau \sum (A_{x,i}-A_{x+j,i}+A_{x+i,j}-A_{x,j})^2.
$$
Let us decompose $\Phi_{x,i}$ into radial and phase variables as follows,
\begin{equation}
\Phi_{x,i}=(\sqrt{\rho_0}+\delta \rho_{x,i})e^{i\phi_{x,i}}.
\label{rho1}
\end{equation}
By substituting Eq.(\ref{rho1}) into Eq.(\ref{S}), the first term of $S$
becomes
\begin{eqnarray}
S_\tau&=&
\int d\tau \sum_{x,i}\Phi^\ast_{x,i}\partial_\tau \Phi_{x,i}\nonumber \\
&=&\int d\tau \sum_{x,i}2i \dot{\phi}_{x,i}\sqrt{\rho_0}\delta 
\rho_{x,i}+O((\delta \rho_{x,i})^2),
\label{Phi2}
\end{eqnarray}
where we have used the periodic boundary condition on $\delta \rho_{x,i}$
and $\phi_{x,i}$ (mod $2\pi$) in the $\tau$-direction, 
and $\dot{\phi}_{x,i}=\partial_\tau \phi_{x,i}$.
Similarly the terms in $H$ Eq.(\ref{H}) are expressed as 
\begin{eqnarray}
H_t&\equiv& 
-t\sum_{x,i\neq j}(\Phi_{x,i}U^\ast_{x+i,j}\Phi^\ast_{x+j,i}U^\ast_{x,j}
+ \mbox{c.c.})  \nonumber  \\
&=&-2t\rho_0 \sum_{x,i\neq j}\cos (\phi_{x,i}
-\phi_{x+i,j}-A_{x,j}-A_{x+i,j})+O(\delta \rho_{x,i}), \nonumber  \\
H_C&\equiv& -C\sum_{x,i\neq j}(\Phi_{x,i}\Phi^\ast_{x+i,j}\Phi_{x+j,i}
\Phi^\ast_{x,j}+ \mbox{c.c.}) \nonumber \\
&=&-2C\rho_0^2 \sum_{x,i\neq j}\cos (\phi_{x,i}-\phi_{x+i,j}
+\phi_{x+j,i}-\phi_{x,j})+O(\delta \rho_{x,i}),  \nonumber  \\
H_D&\equiv&
D\sum_{x,i\neq j}(\Phi_{x,i}\Phi^\ast_{x+i,j}U^\ast_{x+j,i}U^\ast_{x,j}
+ \mbox{c.c.}) \nonumber \\
&=&2D\rho_0^2 \sum_{x,i\neq j}\cos (\phi_{x,i}-
\phi_{x+i,j}-A_{x+j,i}-A_{x,j})+O(\delta \rho_{x,i}),\nonumber \\
H_N&\equiv&
\sum_{x,y,i,j}{\cal V}_{x,y}(N_{x,i}-\rho_0)(N_{y,j}-\rho_0)\nonumber \\
&=&\sum_{x,y,i,j}4{\cal V}_{x,y}\rho_0\delta \rho_{x,i}
\delta \rho_{y,j}+O((\delta \rho_{x,i})^3).
\label{S1}
\end{eqnarray}

From Eqs.(\ref{Phi2}) and (\ref{S1}), we can integrate over $\delta \rho_{x,i}$
in Eq.(\ref{Z}) and obtain, 
\begin{eqnarray}
e^{-S_{\phi}}&\equiv&
\int [d\delta \rho]e^{-S_\tau-\int d\tau H_N} \nonumber \\
&=&e^{-\int d\tau \sum_{x,i,j}
{\cal V}^{-1}_{x,y}\dot{\phi}_{x,i}\dot{\phi}_{y,j}}.
\label{int1}
\end{eqnarray}
Furthermore, we can prove the following identity by straightforward 
calculation (see Appendix A),
\begin{equation}
e^{-S_{\phi}}=\int [dA_0]e^{-\int d\tau H_A},
\label{SAtext}
\end{equation}
with
\begin{equation}
H_A=\int dk \; d^2[e^{-2}\tilde{V}(k)-1]^{-1}
\tilde{A}_0(-k)\tilde{A}_0(k) 
+{1\over e^2}\sum_{x,i,j}(\dot{\phi}_{x,i}-eA_{x,0})
(\dot{\phi}_{x,j}-eA_{x,0}),
\label{HAtext}
\end{equation}
where $\tilde{V}(k)$ is the Fourier transformation of ${\cal V}_{x,y}$ and
the scalar potential $A_{x,0}$ is the field that mediates the
Coulomb interaction (\ref{V}).
($\tilde{A}_0(k)$ is the Fourier transformation of $A_{x,0}$.)

Similarly we can show (see Appendix A),
\begin{equation}
e^{-\int d\tau H_t}=\int [dV^\ast dV]e^{-\int d\tau H_V},
\label{HPsitext}
\end{equation}
where 
\begin{equation}
H_V={1\over 2}\sum_{x,i\neq j}\Big[J^{-1}e^{iA}V^\ast_{x,i}
V_{x+j,i}-V_{x,i}e^{-i\phi_{x,i}}+\mbox{c.c.}\Big],
\label{Psi2text}
\end{equation} 
with $A\equiv A_{x+i,j}+A_{x,j}$ and $J=2t\rho_0$.
As $V_{x+j,i} \propto e^{-iA+i\phi_{x,i}}$, the field $V_{x,i}$ can be
regarded as the Cooper-pair field.

Finally we shall integrate over $\phi_{x,i}$.
To this end, we consider the following integral
\begin{eqnarray}
e^{-\int d\tau H'_{V}}&=&\int [d\phi]\exp\Bigg( \int d\tau 
\Big[-{1\over e^2}\sum_{x,i,j}(\dot{\phi}_{x,i}-eA_{x,0})
(\dot{\phi}_{x,j}-eA_{x,0})  \nonumber \\ 
&&+(d-1)\sum_{x,i}(V_{x,i}e^{-i\phi_{x,i}}+\mbox{c.c.})\Big]\Bigg).
\label{Psi22}
\end{eqnarray}
From Eq.(\ref{Psi22}), it is obvious that
requirement of the invariance of $H'_{V}$ under a 
``local gauge transformation",
\begin{equation}
\phi_{x,i}\rightarrow \phi_{x,i}+e\alpha_{x}(\tau),\;\;
A_{x,0}\rightarrow A_{x,0}+\dot{\alpha}_{x}(\tau), \;\;
V_{x,i}\rightarrow e^{ie\alpha_{x}(\tau)}V_{x,i},
\label{gtrf1}
\end{equation}
determines dependence on $A_{x,0}$ in $H'_{V}$.
Then by simply putting $A_{x,0}=0$, we define 
\begin{equation}
\langle {\cal O} \rangle_0=Z^{-1}_{\phi}\int [d\phi]\;
e^{-{1\over e^2}\int d\tau \sum \dot{\phi}_{x,i}\dot{\phi}_{x,j}}
{\cal O}, \;\;
Z_{\phi}=\int [d\phi]\;
e^{-{1\over e^2}\int d\tau \sum \dot{\phi}_{x,i}\dot{\phi}_{x,j}}.
\label{Z0}
\end{equation}
It is not difficult to show,\footnote{It should be remarked here that
the Green function like 
$\langle e^{i(\phi_{x,i}(\tau_1)-\phi_{x,i}(\tau_2))}\rangle_0$, which is 
invariant under a $\tau$-independent ``local gauge transformation" 
$\phi_{x,i}(\tau)\rightarrow \phi_{x,i}(\tau)+\beta_{x,i}$, is well-defined,
whereas $\langle \phi_{x,i}(\tau_1)\phi_{y,i}(\tau_2)\rangle_0$ is not.}
\begin{equation}
\langle \phi_{x,i}(\tau_1)\phi_{y,i}(\tau_2)\rangle_0
=\delta_{xy}\; \mbox{lim}_{\epsilon \rightarrow 0}
{e^2\over \epsilon}e^{-\epsilon |\tau_1-\tau_2|},
\label{phi2}
\end{equation}
where $\epsilon$ is the infrared cutoff.
Then, 
\begin{equation}
\langle e^{i(\phi_{x,i}(\tau_1)-\phi_{x,i}(\tau_2))}\rangle_0=
\exp [-e^2|\tau_1-\tau_2|].
\label{Green1}
\end{equation}
From Eqs.(\ref{Psi22}) and (\ref{Green1}),
the lowest-order terms of the cumulant expansion are obtained as 
\begin{eqnarray}
-H'_{V}|_{A_0=0}&=&(d-1)^2 \Big\langle \int d\tau_2 \sum_{x,i}
V_{x,i}(\tau)V_{x,i}^\ast(\tau_2)e^{-i(\phi_{x,i}(\tau)-
\phi_{x,i}(\tau_2))}\Big\rangle_0  \nonumber \\
&=&(d-1)^2\int d\tau_2V_{x,i}(\tau)V_{x,i}^\ast(\tau_2)
e^{-e^2 |\tau-\tau_2|} \nonumber \\
&=& \lambda_1 \sum_{x,i}|V_{x,i}(\tau)|^2
-{\lambda_2 \over 2} \sum_{x,i}|\dot{V}_{x,i}(\tau)|^2+\cdots,
\label{PsiPsi}
\end{eqnarray}
where 
\begin{eqnarray}
&& \lambda_1=(d-1)^2\int d\tau \; e^{-e^2 |\tau|}=(d-1)^2
{2\over e^2}, \nonumber \\
&& \lambda_2=(d-1)^2 \int d\tau \tau^2 e^{-e^2 |\tau|}=(d-1)^2{4\over e^6}.
\label{alpha}
\end{eqnarray}

From Eq.(\ref{Psi2text}), the spatial coupling of $V_{x,i}$ in the 
Hamiltonian is given as (neglecting the gauge field for simplicity)
$H_V\sim V^\ast_{x,i}V_{x+j,i}+\mbox{c.c.}$, and therefore
we replace as $V_{x,i}\rightarrow (-)^{|x|}\sqrt{J}V_{x,i}$ 
in the action where $|x|=1(-1)$ for even (odd) site.
After this replacement, the first term of $H_V$ becomes
\begin{equation}
-{1 \over 2}\sum_{x,i\neq j}\Big[e^{iA}V^\ast_{x,i}V_{x+j,i}+
\mbox{c.c.}\Big] 
={1 \over 2}\sum_{x,i\neq j}|D_jV_{x,i}|^2
-(d-1)\sum_{x,i} |V_{x,i}|^2,
\label{HPsi2}
\end{equation} 
where $D_jV_{x,i}=U_{x,j}U_{x+i,j}V_{x+j,i}-V_{x,i}$.

Let us define the effective action $S_{\rm CP}$ of the Cooper-pair field 
$V_{x,i}$ and the gauge field as,
\begin{equation}
Z=\int [dV dA]e^{-S_{\rm CP}-S_{\rm em}}=
\int [d\Phi dA]e^{-S}.
\label{ZSe}
\end{equation}
Then up to the second-order terms,
\begin{eqnarray}
S_{\rm CP}&\sim &\int d\tau \Bigg[
{1 \over 2}\sum_{x,i\neq j}|D_jV_{x,i}|^2 
+{J\lambda_2 \over 2} \sum_{x,i}|D_\tau{V}_{x,i}(\tau)|^2 \nonumber \\
&& -(J\lambda_1+d-1)\sum_{x,i} |V_{x,i}|^2 
 +\int dk \; d^2[e^{-2}\tilde{V}(k)-1]^{-1}\tilde{A}_0(-k)\tilde{A}_0(k)
\Bigg],
\label{S2}
\end{eqnarray}
where $D_{\tau}V_{x,i}=(\partial_{\tau}-ieA_{x,0})V_{x,i}$.

At long distances, the last term of $S_{\rm  CP}$ in Eq.(\ref{S2})
becomes the usual relativistic kinetic terms of $A_{x,0}$,
\begin{equation}
\int d\tau\int dk\; {k}^2\tilde{A}_0(-{k},\tau)\tilde{A}_0({k},\tau),
\label{relA0}
\end{equation}
whereas the effects of the short-range part of ${\cal V}_{x,y}$
generate terms like $|V_{x,i}|^4$. Similar quartic terms also
appear in the higher-order terms of the cumulant expansion. 
Furthermore, the second and third terms of $H$ in Eq.(\ref{H}) 
are expressed in terms of $V_{x,i}$ by the perturbative calculation
in powers of $C,\;D$ in the cumulant expansion.

In the derivation of the GL model action $S_{\rm CP}$ given in this 
section, only the lowest-order terms of $\delta\rho_{x,i}$ are considered.
It is obvious that higher-order terms like $O((\delta \rho_{x,i})^2)$-terms
in Eq.(\ref{Phi2}) give nonvanishing contributions to the GL model.
However we think that the full GL model, which contains all relevant
terms for studying the SC phase transition, has a similar form to that
obtained in the above.
The expansion in powers of the order-parameter field $V_{x,i}$,
the invariance under a $\pi/2$-rotation in the spatial planes and
the invariance under a local gauge transformation,
\begin{equation}
U_{x,i} \rightarrow e^{i\omega_{x+i}}U_{x,i}e^{-i\omega_x}, \;\;\;
V_{x,i} \rightarrow e^{-i\omega_{x+i}}V_{x,i}e^{-i\omega_x},
\label{localGT2}
\end{equation}
which originates from Eq.(\ref{localGT}),
restrict the form of the GL action.
Effective field-theory model for the granular $s$-wave SC has been
derived in a similar way\cite{GL1}.
Then in order to study phase structure, order of phase transitions, etc, 
we propose action of the GL model as follows, which is a kind of the 
U-V model proposed previously,
\begin{eqnarray}
S_{\rm GL} &=&g\Big[ c_{1} \sum_{x,\mu \neq \nu}  F^2_{\mu\nu}(x) 
 - c_{2} \sum_{x,\mu \neq i} ( U_{x,\mu} V_{x+\mu,i} 
U_{x+\nu,\mu} V_{x,i}^{\ast} + C.C. ) \nonumber \\
&& - d_{2} \sum_{x,i \neq j} ( U_{x,i} U_{x+i,j} 
V_{x+j,i} V^{\ast}_{x,j} + C.C. )  
 - c_{3} \sum_{x,i \neq j} ( V_{x,i} V^{\ast}_{x+i,j} 
V_{x+j,i} V^{\ast}_{x,j} + C.C. )\Big] \nonumber \\
&& + \alpha \sum_{x,i} \left| V_{x,i} \right|^{4} - \beta \sum_{x,i} 
\left| V_{x,i} \right|^{2},
\label{action}
\end{eqnarray}
where $F_{\mu\nu}(x)=A_{x,\mu}+A_{x+\mu,\nu}-A_{x,\nu}-A_{x+\nu,\mu}$
($\mu,\nu=0,1,...,d$) and we have introduced a lattice also
in the $\tau$-direction.
It should be noticed that the gauge field $U_{x,\mu}\; (A_{x,\mu})$ also has 
the $\tau$-component $U_{x,0}=e^{iA_{x,0}}$ besides the spatial 
ones\footnote{It should be noticed that in the action (\ref{action})
$V_{x,i}$ couples with both $U_{x,0}$ and $U_{x+i,0}$.
This means that $V_{x,i}$ has electric charge at $x$ and $x+i$ as the
original Cooper pair given by Eq.(\ref{C-P}) does.},
whereas the Cooper-pair field $V_{x,i}$ has only spatial components.
For numerical calculation, we fix values of the parameters 
$c_1, \cdots, \beta$ in
Eq.(\ref{action}) and study phase structure by varying $g$.
For large $g$, the electromagnetic interactions between $V_{x,i}$
is getting weak, and the hopping of $V_{x,i}$ is enhanced.
Because of the uncertainty relation between the number $n_{x,i}$ and 
phase $\varphi_{x,i}$, $\Delta n_{x,i}\cdot \Delta\varphi_{x,i} \ge 1$, 
a SC state, in which $V_{x,i}$'s are stabilized, is formed.
For small $g$, on the other hand, charge density $n_{x,i}$ tends to have
definite distributions and as a result an insulating Wigner crystal or 
the normal metallic state appears.
In other words, the parameter $g$ plays a role of $1/\hbar$

%%%%%%%%%%%%%%%%%%%%%%%%%%%%%%%%%%%%%%%%%%%%%%%%%%%%%%%%%%%%%%%%%%
\setcounter{equation}{0}
\section{Numerical results in (2+1)D}

Let us turn to the numerical studies of the present lattice U-V
model for the SC phase transition in two and three spatial 
dimensions\cite{numerical}.
In order to investigate phase structure, we measured the 
``internal energy" $E$ and the ``specific heat" $C$,
\begin{equation}
E\equiv -\langle S_{\rm GL} \rangle/V, \;\; \; 
C\equiv \langle (S_{\rm GL}-\langle S_{\rm GL} \rangle)^2 \rangle/V,
\label{EC}
\end{equation}
where $V=L^3$ is the system size.
For investigation on 2D thin film SC materials,
we consider 3D symmetric lattice as we are studying the QPT at $T=0$.
In the action $S_{\rm GL}$ in Eq.(\ref{action}), we fix the values of 
$c_1, \cdots, \beta$ and measured $E$ and $C$ as a function of $g$.
As we explained in the previous section, the parameter $g$ controls
quantum fluctuations, and therefore an observed phase transition is 
nothing but a QPT.

\begin{figure}[htbp]
\begin{center}
%\leavevmode
\epsfxsize=6cm
\epsffile{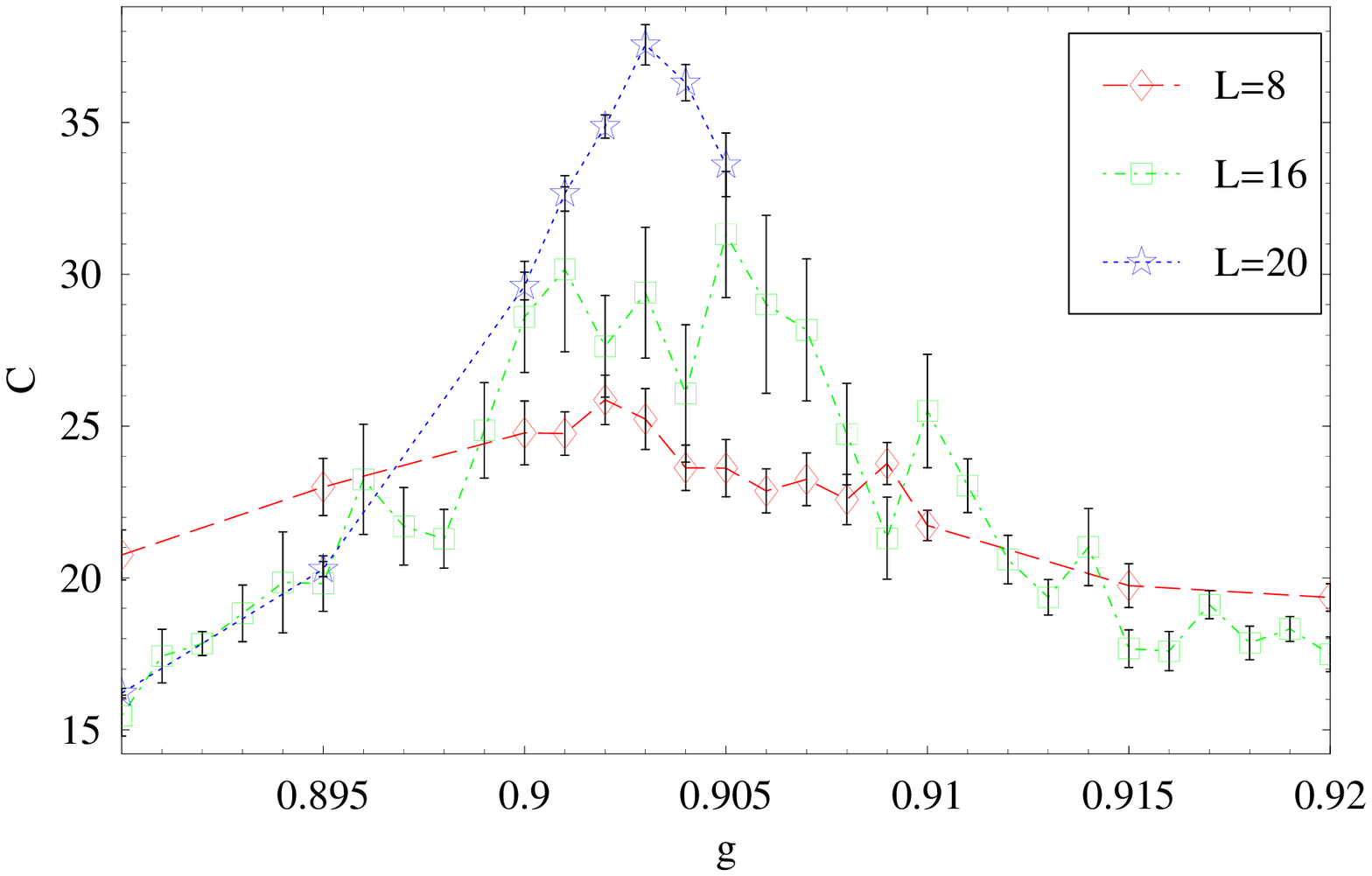}
\epsfxsize=6cm
\epsffile{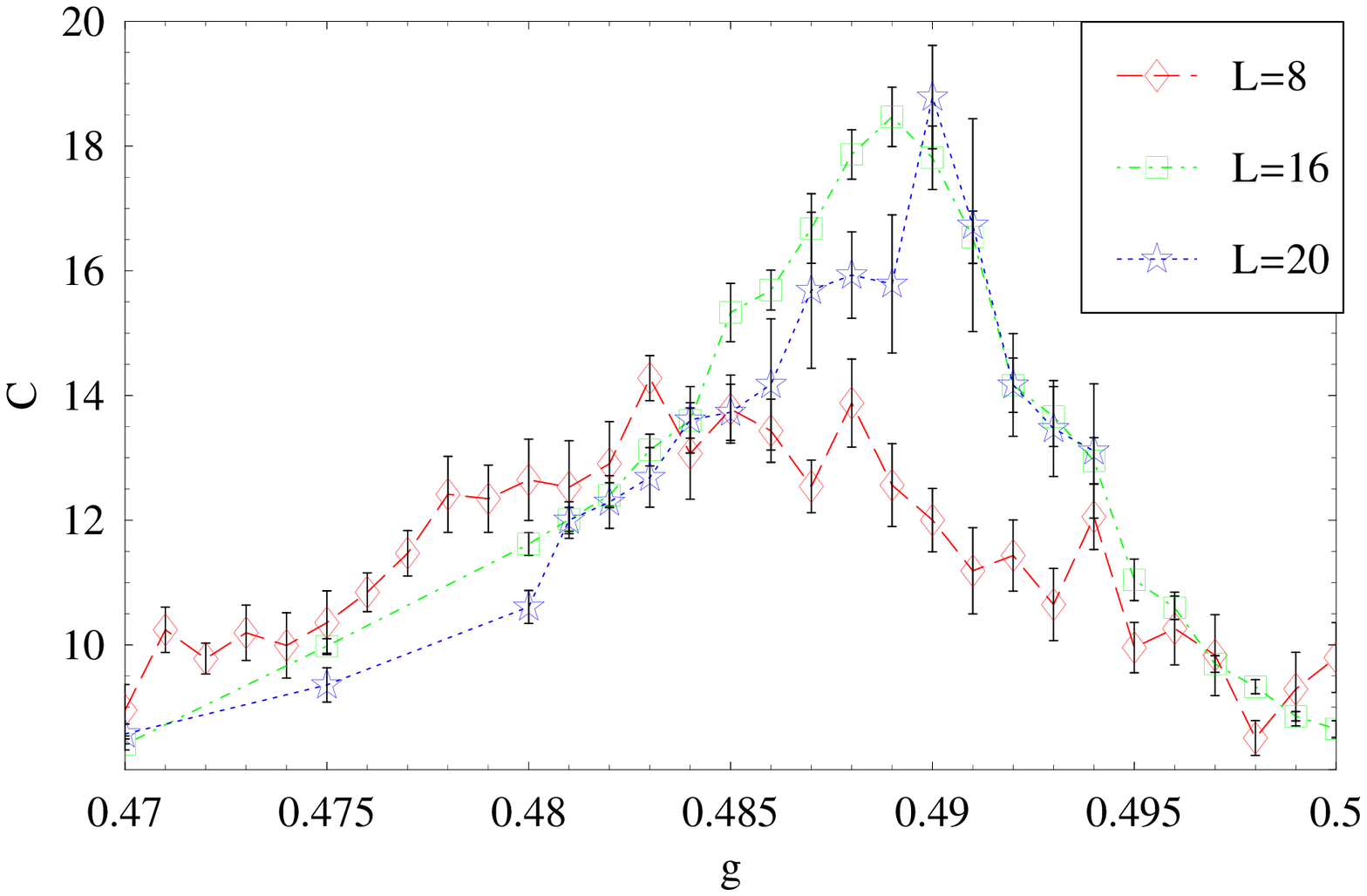}
\caption{System-size dependence of the ``specific heat" $C$ for 
$\alpha=\beta=5$ (left) and the London limit (right) of the $(2+1)$D UV model. 
The peak develops as $L$
is getting larger in each case. The result indicates that the 
second-order phase transition occurs at around $g=0.902 (0.490)$
for the $\alpha=\beta=5$ (London limit) case.
}
\label{figC1}
\end{center}
\end{figure}

We mostly studied the cases with the parameters $c_1=c_2=c_3=-d_2=1$,
$\alpha=\beta=3,5,10$ and also the London limit of $V_{x,i}$.
The negative value of $d_2$ means that the configurations 
corresponding to the $d_{x^2-y^2}$-wave SC are enhanced for large $g$.
In all these cases, we have observed no anomalous behavior of $E$,
whereas we found that $C$ exhibits a sharp peak as $g$ is varied.
We show the system-size dependence of $C$ in Fig.\ref{figC1} 
for the cases of $\alpha=\beta=5$ and the London limit.
It is obvious that the peak develops as $L$ is getting larger in both cases.
Similar behavior of $C$ was observed for the case with $\alpha=\beta=3,\; 10$.
Then we conclude that a second-order phase transition takes place as $g$ is
increased.

%---------------------------------------------------
%----------------------------------------------------

\begin{figure}[htbp]
\begin{center}
%\leavevmode
\epsfxsize=6cm
\epsffile{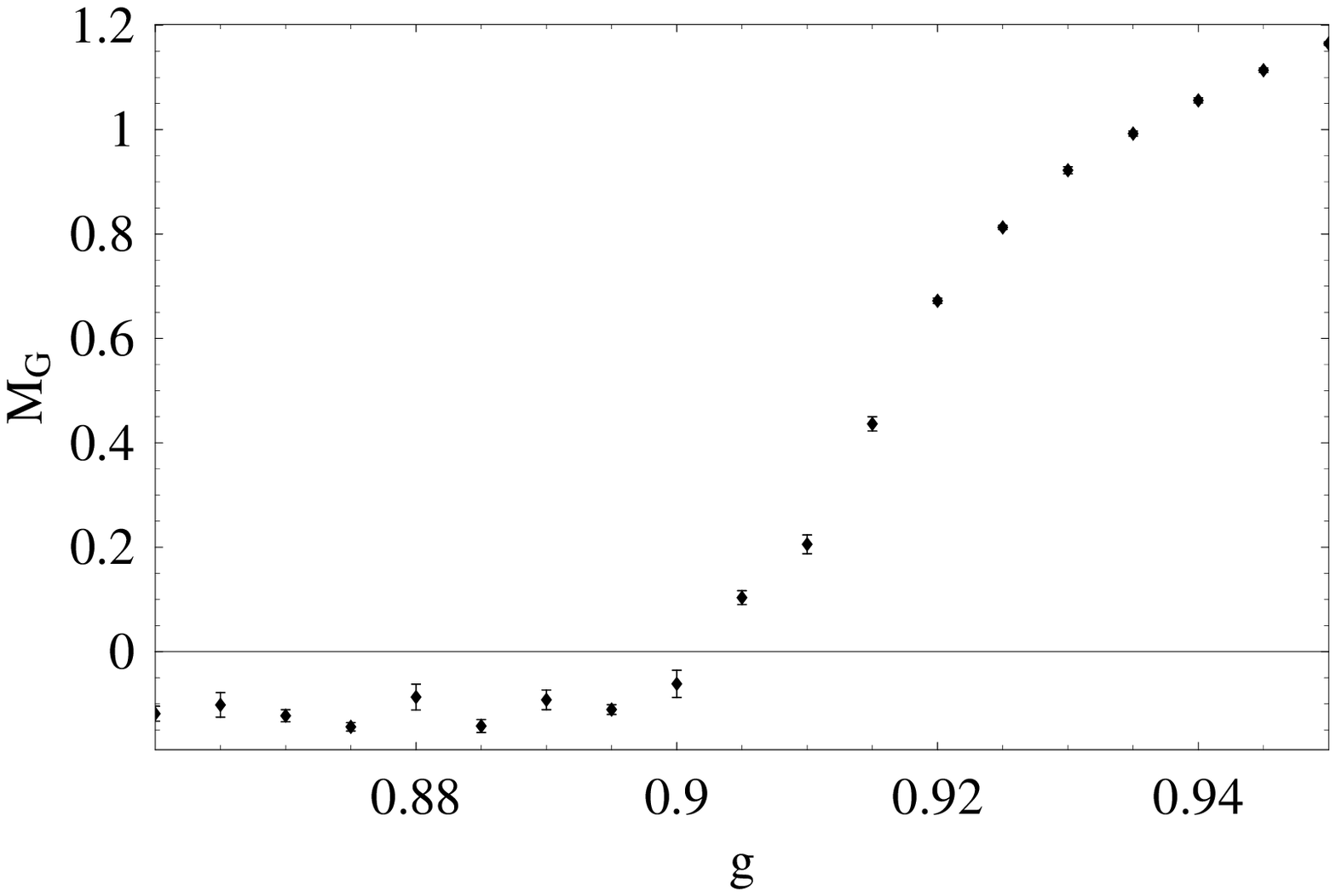}
\epsfxsize=6cm
\epsffile{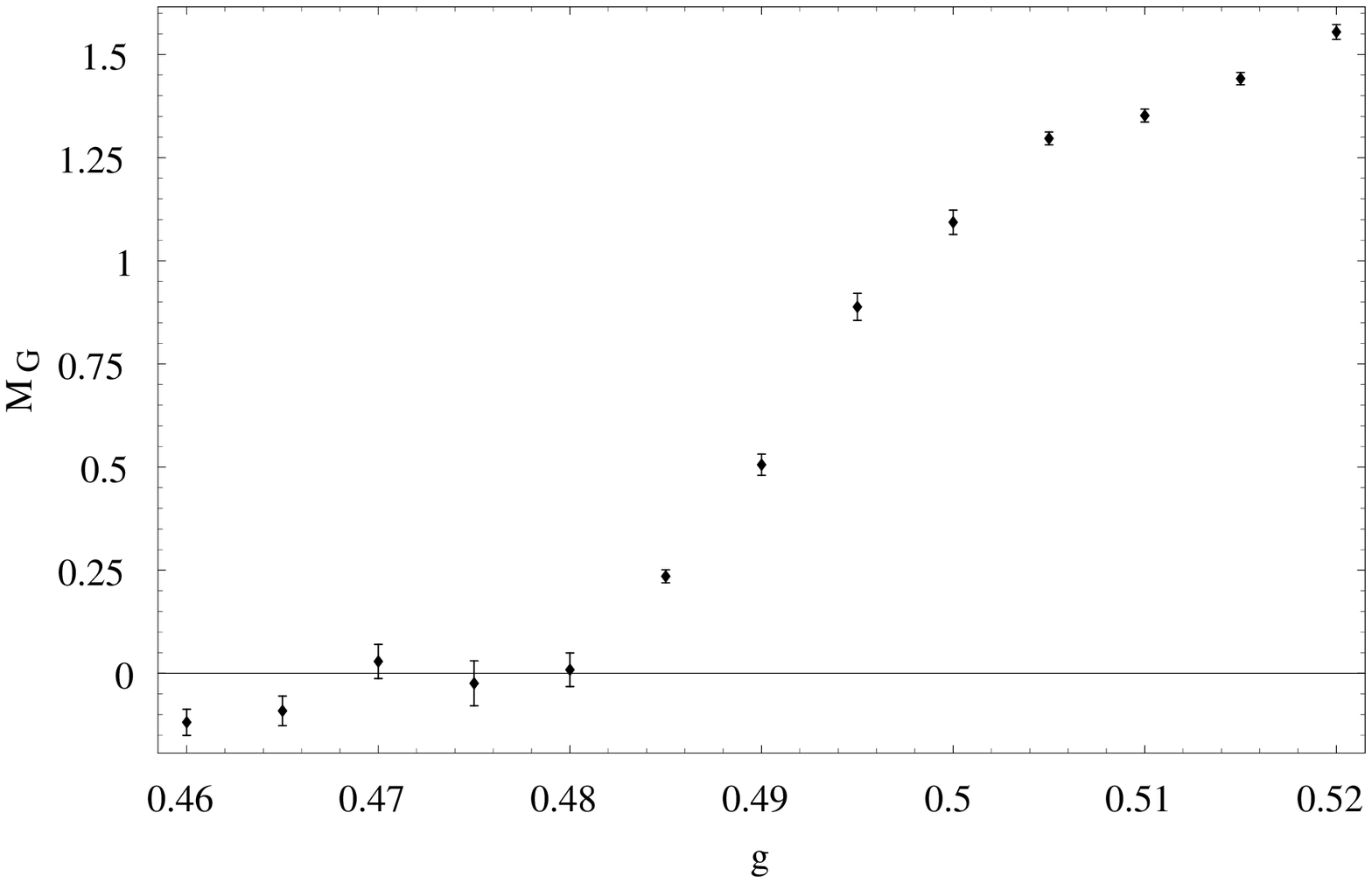}
\caption{Gauge-boson mass $M_G$ for the cases of $\alpha=\beta=5$ (left) and 
London limit (right) of the $(2+1)$D UV model.
$M_G$ develops at the critical coupling obtained by the measurement of $C$.
This result indicates that the SC phase transition occurs and the order
of the transition is of second order.
}
\label{figMG1}
\end{center}
\end{figure}

In order to verify that the observed phase transition is a SC transition,
we measured gauge-invariant gauge-boson mass $M_G$ (the inverse of the magnetic
penetration depth).
$M_G$ is defined as the inverse of the correlation length of the operator
$O(x)=\sum_{i,j=1,2}\epsilon_{ij}\sin (F_{ij}(x))$, where 
$\epsilon_{12}=-\epsilon_{21}=1$\cite{MG}.
In order to evaluate $M_G$ accurately, we introduce Fourier transformed
operator of $O(x)$ in the 1-2 plane,
\begin{equation}
\tilde{O}(x_0)=\sum_{x_1,x_2}O(x)e^{ip_1x_1+ip_2x_2}.
\label{Ftrf1}
\end{equation}
Then we expect the following behavior,
\begin{equation}
\langle \tilde{O}(x_0+\tau)\tilde{O}(x_0)\rangle \propto
e^{-\sqrt{p^2_1+p^2_2+M^2_G}\tau}.
\label{correlationF}
\end{equation}
In the numerical calculation, we put $p_1=p_2={1 \over L}$ and 
verified that the correlator $\langle \tilde{O}(x_0+\tau)\tilde{O}(x_0)\rangle$
exhibits the exponential decay as in Eq.(\ref{correlationF}).
More precisely, we define $M_G$ as $M_G={\rm sign} (\lambda^2-\vec{p}^2)
\sqrt{\lambda^2-\vec{p}^2}$, where $\lambda$ is the inverse correlation
length of the correlator of $\tilde{O}(x_0)$.
In Fig.\ref{figMG1}, we show the calculations of $M_G$ for $\alpha=\beta=5$ and
the London limit cases.
The negative value of $M_G$ in Fig.\ref{figMG1} is the finite-size 
effect\cite{MG}.
The results show that in both cases $M_G$ develops continuously from the phase
transition point obtained by the measurement of $C$.
At first-order phase transition points, the $M_G$ exhibits a sharp
jump from vanishing to a finite value\cite{UV2}.
Therefore we conclude that the phase transition observed by the measurement 
of $C$ is a second-order SC transition.
We have observed similar behavior of $M_G$ for the other cases with 
$\alpha=\beta=3,\; 10$.

%---------------------------------------------------
%----------------------------------------------------

%%%%%%%%%%%%%%%%%%%%%%%%%%%%%%%%%%%%%%%%%%%%%%%%%%%%%%%%%%%%%%%%%%
\setcounter{equation}{0}
\section{Numerical results in (3+1)D}

In this section, we shall show results of the numerical study of the 
(3+1)D cases.
As most of the materials of the unconventional SC have quasi-2D structure,
studies on 2D and 3D systems are useful to obtain the physical
picture of the real materials.
It is known that quasi-2D systems except for thin film samples 
exhibit 3D properties in critical regions.

\begin{figure}[htbp]
\begin{center}
%\leavevmode
\epsfxsize=8cm
\epsffile{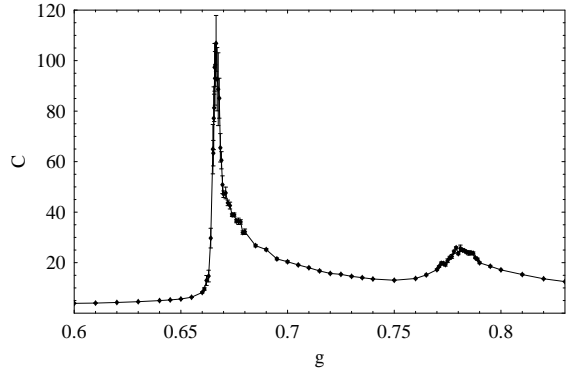}
\caption{Specific heat $C$ for $(3+1)$D U-V model with $\alpha=\beta=5$.
System size is $8^4$. There exist two peaks at $g\sim 0.67$ and $0.78$.
}
\label{figC2}
\end{center}
\end{figure}

\begin{figure}[htbp]
\begin{center}
\epsfxsize=7cm
\epsffile{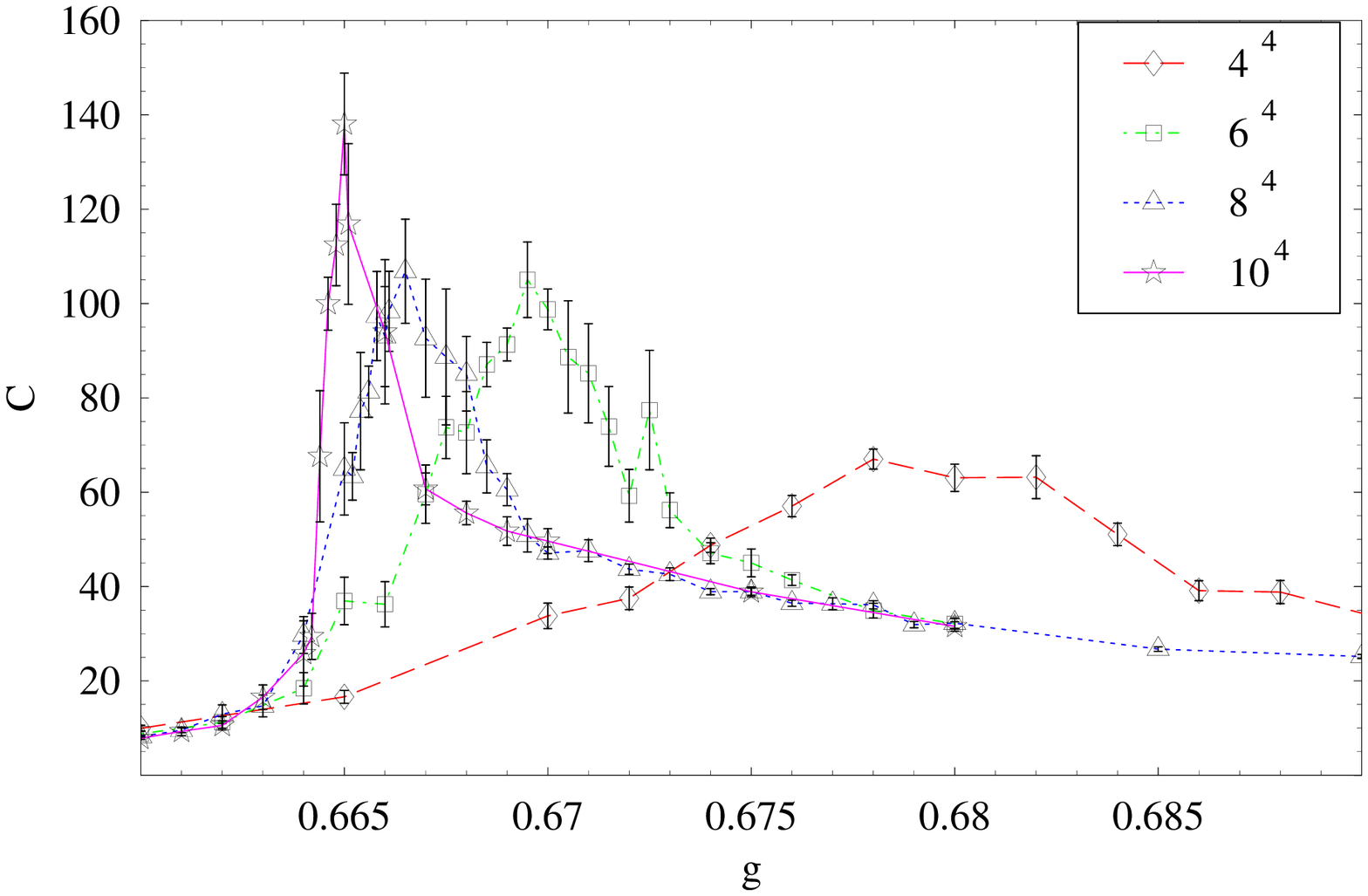}
\epsfxsize=7cm
\epsffile{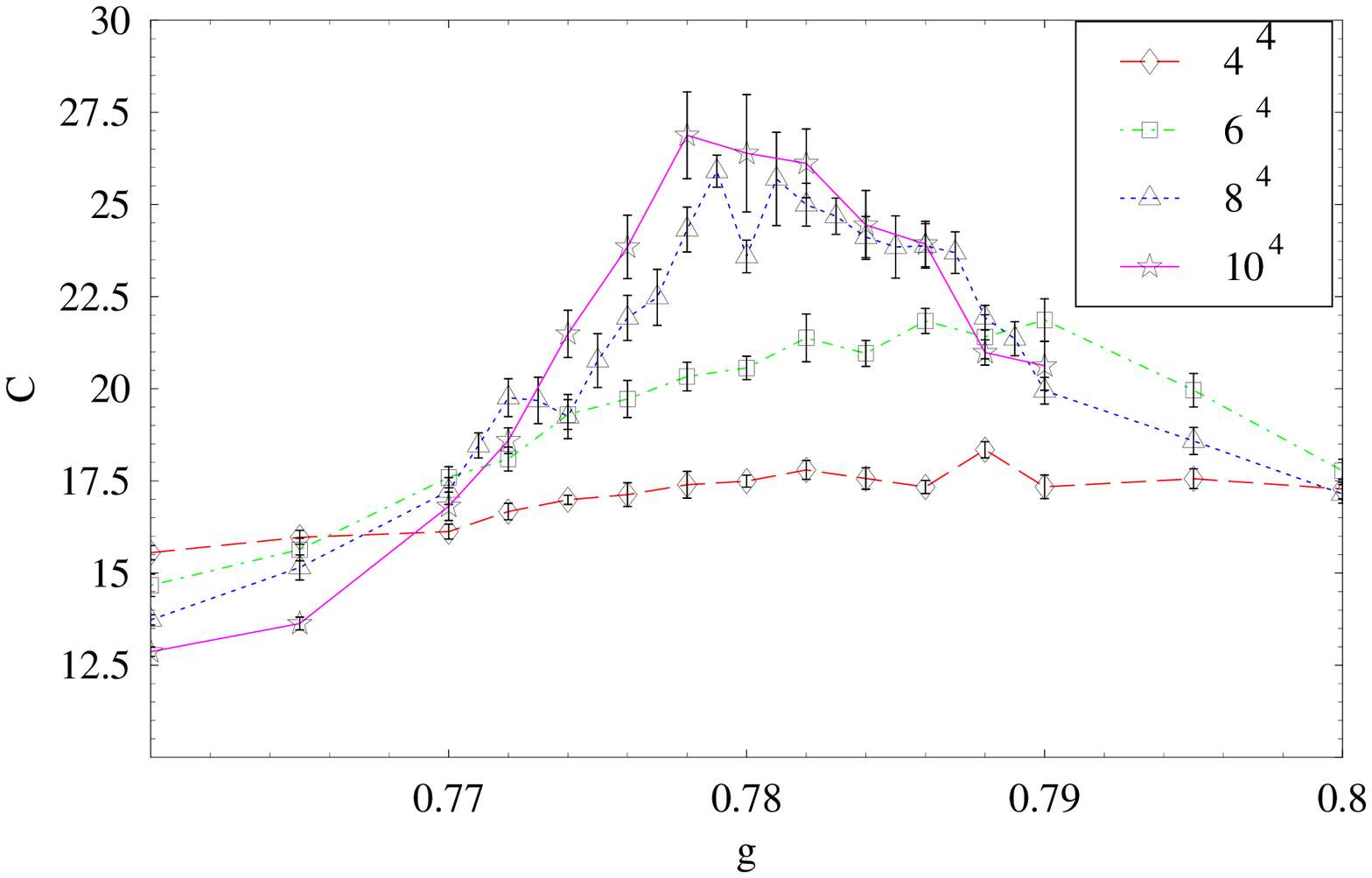}
\caption{System-size dependence of two peaks in $C$ for $\alpha=\beta=5$.
Result indicates that both peaks are signal of second-order phase transition.
}
\label{figC3}
\end{center}
\end{figure}

\begin{figure}[htbp]
\begin{center}
\epsfxsize=8cm
\epsffile{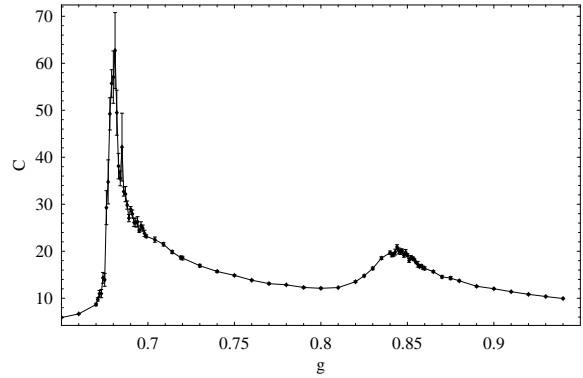}
\caption{Specific heat $C$ for $\alpha=\beta=10$ with system size $8^4$.
There are two peaks.
}
\label{figC4_1}
\end{center}
\end{figure}

\begin{figure}[htbp]
\begin{center}
\epsfxsize=7cm
\epsffile{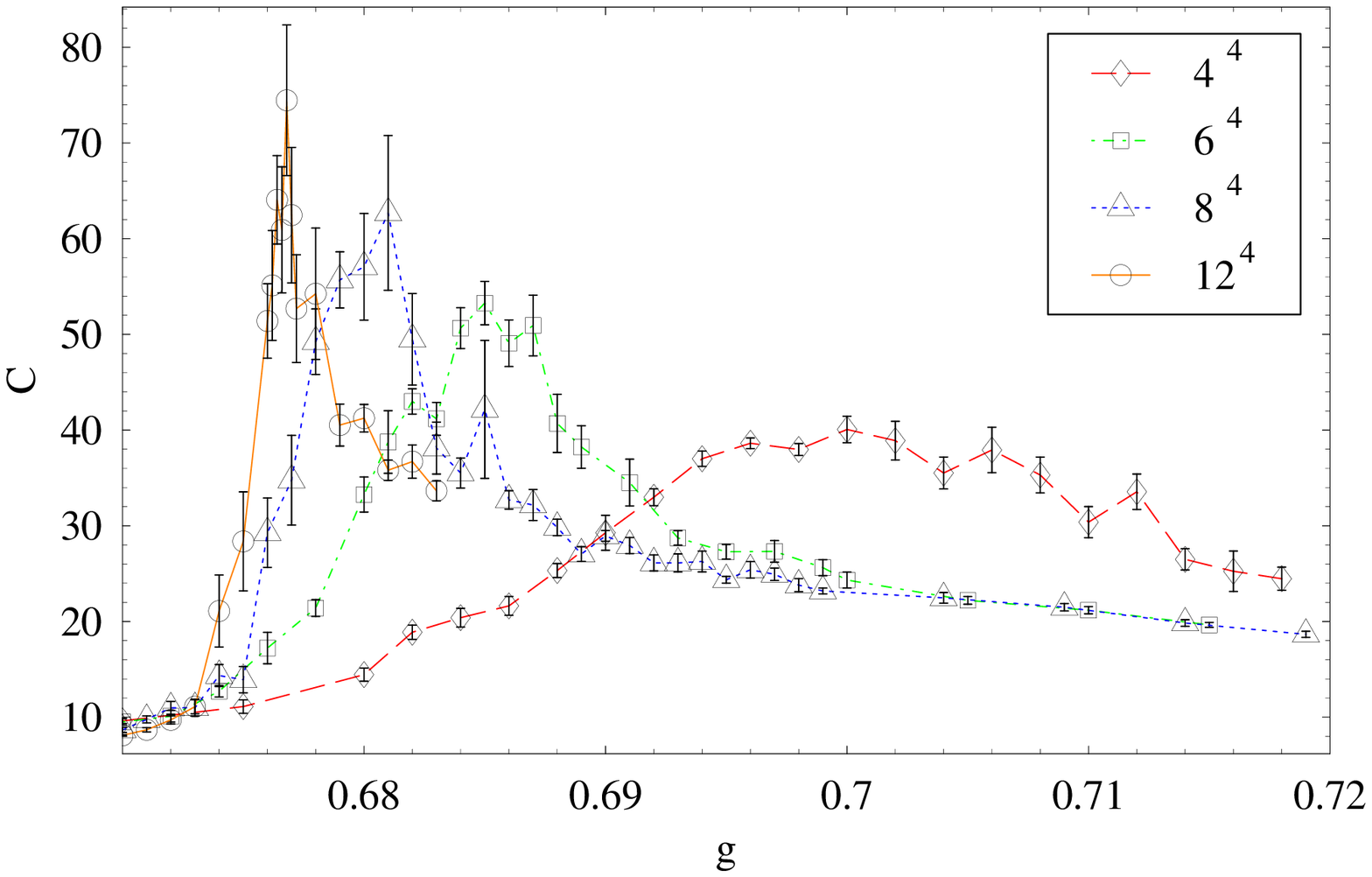}
\epsfxsize=7cm
\epsffile{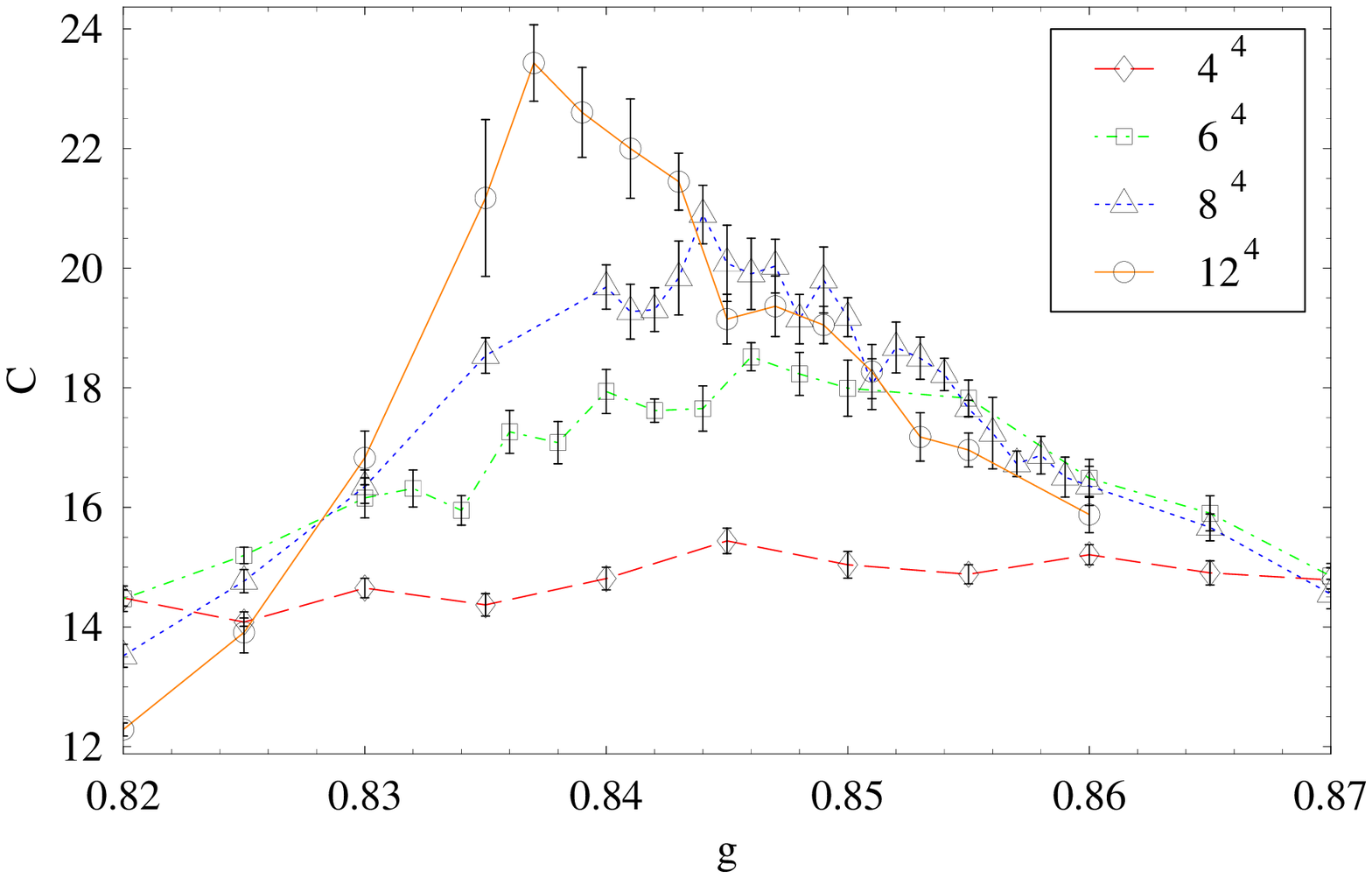}
\caption{System size dependence of two peaks in $C$ for $\alpha=\beta=10$.
Result indicates that both peaks are signal of second-order phase transition.
}
\label{figC4_2}
\end{center}
\end{figure}

As in the $(2+1)$D case, we first show the calculation of $C$ in
Fig.\ref{figC2} for the case $\alpha=\beta=5$.
There exist two peaks at $g\sim 0.67$ and $0.78$.
Then we carefully studied the system-size dependence of each peak
and the results are shown in Fig.\ref{figC3}.
From the calculations in Fig.\ref{figC3}, we conclude that 
the both peaks are signal of second-order phase transition
as they have the system-size dependence.
Similar behavior of $C$ was observed in the case with 
$\alpha=\beta=10$ (see Figs.\ref{figC4_1} and \ref{figC4_2})
and the London limit.

%---------------------------------------------------
%----------------------------------------------------

%---------------------------------------------------
%----------------------------------------------------

%---------------------------------------------------
%----------------------------------------------------
%---------------------------------------------------
%----------------------------------------------------

\begin{figure}[htbp]
\begin{center}
%\leavevmode
\epsfxsize=7cm
\epsffile{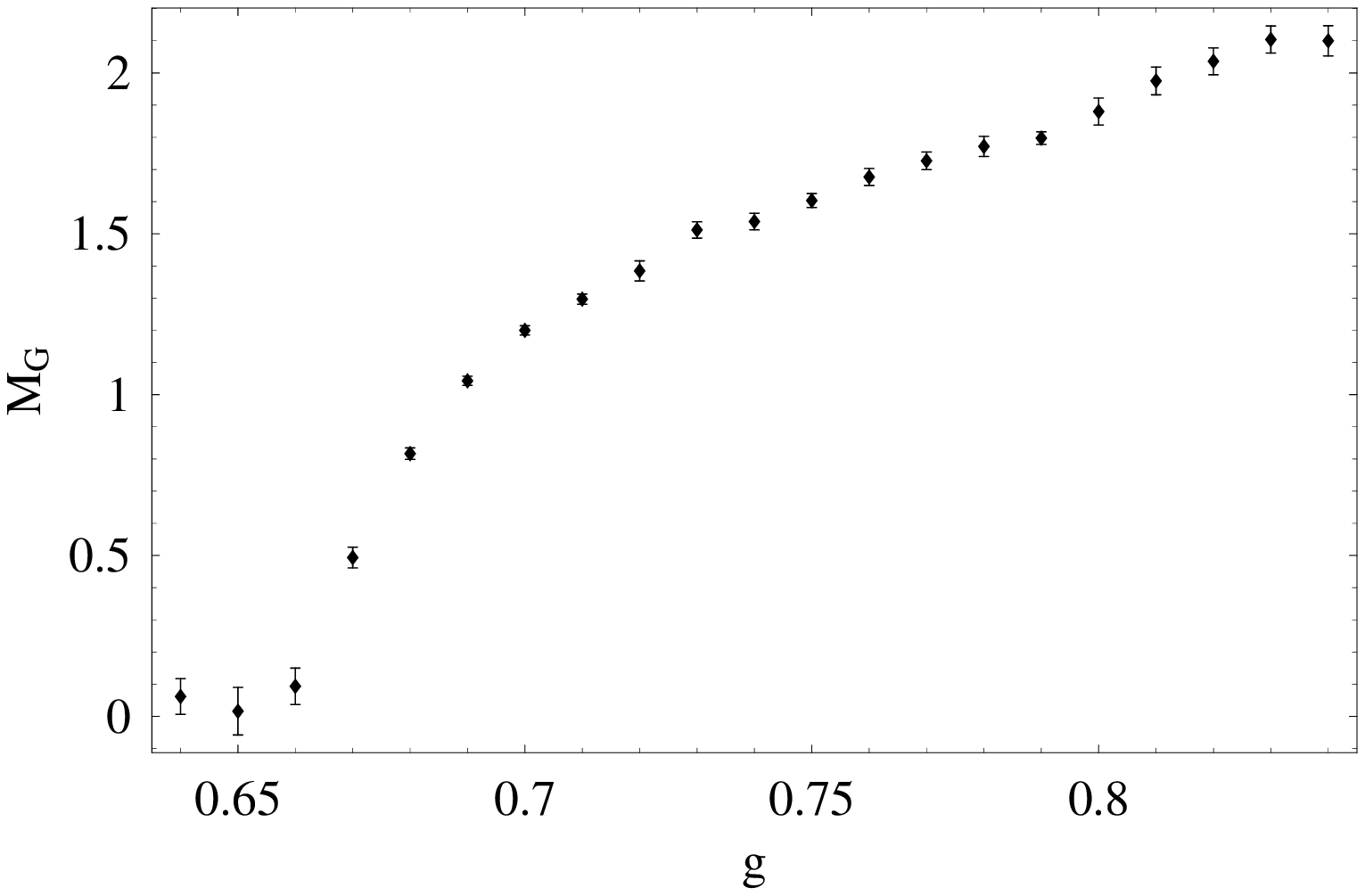}
\caption{Gauge-boson mass $M_G$ in $(3+1)$D system for the 
$\alpha=\beta=5$ 
case. $M_G$ starts to develop at the first critical coupling $g_c\sim 0.67$.
Furthermore at the second critical coupling $g'_c\sim 0.78$, $M_G$ slightly 
changes its behavior upwards.
}
\label{figMG2_1}
\end{center}
\end{figure}

\begin{figure}[htbp]
\begin{center}
%\leavevmode
\epsfxsize=7cm
\epsffile{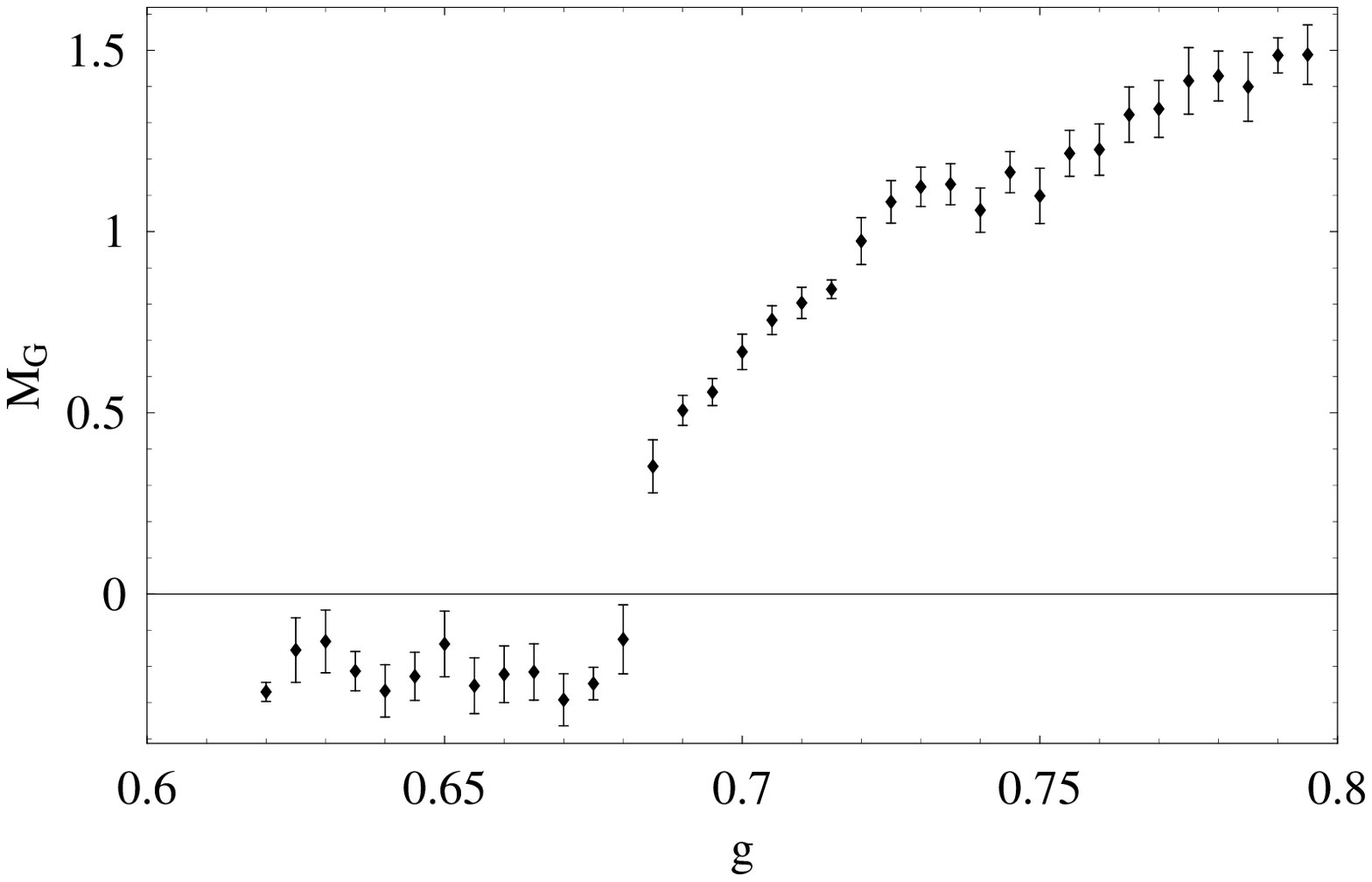}
\epsfxsize=7cm
\epsffile{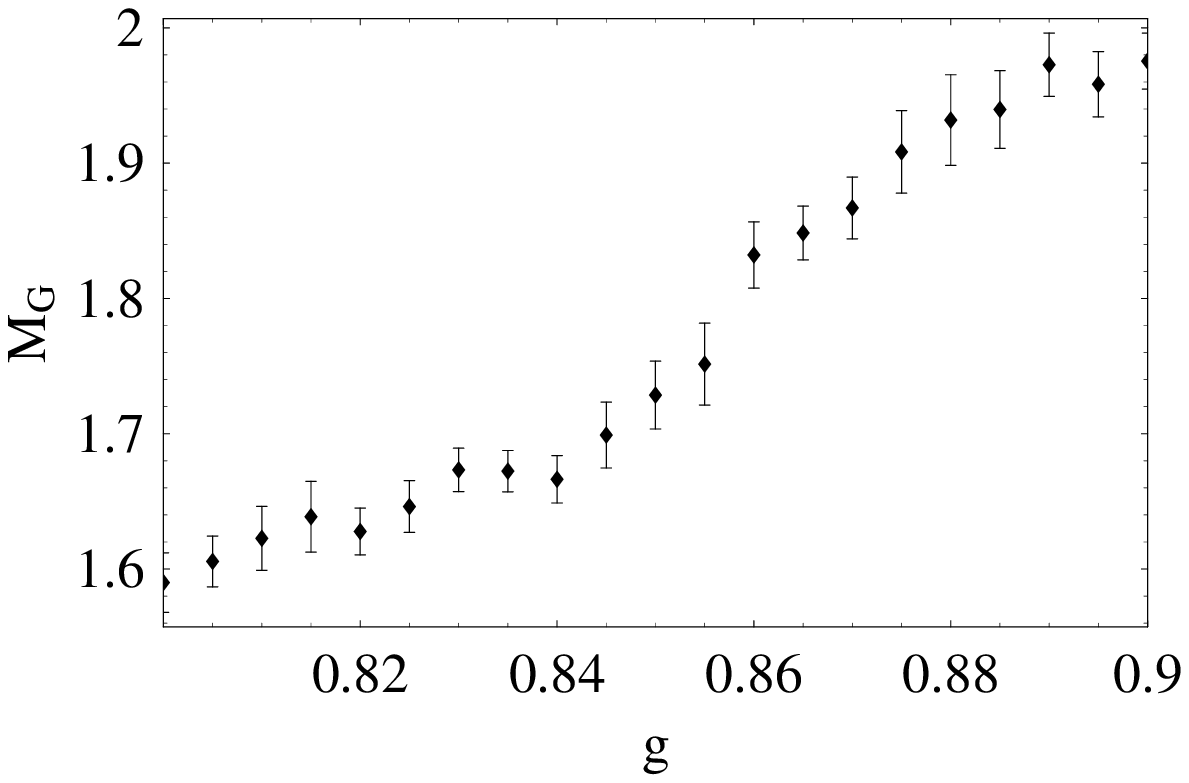}
\caption{Gauge-boson mass $M_G$ in $(3+1)$D system for the 
$\alpha=\beta=10$ 
case. $M_G$ starts to develop at the first critical coupling $g_c\sim 0.68$.
Furthermore at the second critical coupling $g'_c\sim 0.84$, $M_G$ slightly 
changes its behavior upwards.
}
\label{figMG2_2}
\end{center}
\end{figure}

In order to investigate physical meaning of the phase transitions,
we measured the gauge-boson mass $M_G$.
Results are given in Figs.\ref{figMG2_1} and \ref{figMG2_2}.
$M_G$ starts to develop at the first critical coupling $g_c\sim 0.67$.
Furthermore at the second critical coupling $g'_c\sim 0.78$, $M_G$ slightly 
changes its behavior upwards.
Therefore, the SC phase transition occurs at $g_c\sim 0.67$.
On the other hand at $g'_c\sim 0.78$, a certain mechanism suppressing 
fluctuations of the gauge field $A_{x,\mu}$ further starts to work.

%---------------------------------------------------
%----------------------------------------------------

%---------------------------------------------------
%----------------------------------------------------

It is quite helpful to study the monopole density of $U$ and $V$ gauge fields 
for understanding physical meaning of the second phase transition.
In the ordinary $s$-wave SC state, vortices appear as low-energy excitations.
In the GL theory of the $s$-wave SC, the vector potential $\vec{A}^{\rm em}$ 
and the order parameter $\Psi$ are given as follows for vortex 
configurations\cite{vortex},
\begin{equation}
\vec{A}^{\rm em}={1 \over 2e}{\rm grad}\; \theta, \;\; 
\Psi=|\Psi|e^{i\theta}, \;\;
\end{equation}
where 
\begin{equation}
\oint \vec{A}^{\rm em}\cdot d\vec{l}={1 \over 2e}\Delta \theta=
{1\over 2e}2n\pi, \;\; n=\pm 1, \pm 2,\cdots.
\end{equation}
In the present U-V model, similar vortex configurations exist.
However as the $V$-field is defined on links, the $V$-field has also
monopole-like configurations just like the ordinary compact U(1) 
gauge field, i.e., vortices can terminate at monopole and anti-monopole.
The phase transition from the SC to normal states occurs as a result of 
the condensation of (infinitely long) vortices, but we can also expect that
some phase transition, which is related with monopole dynamics, 
occurs at a certain coupling constant.

\begin{figure}[htbp]
\begin{center}
\epsfxsize=7cm
\epsffile{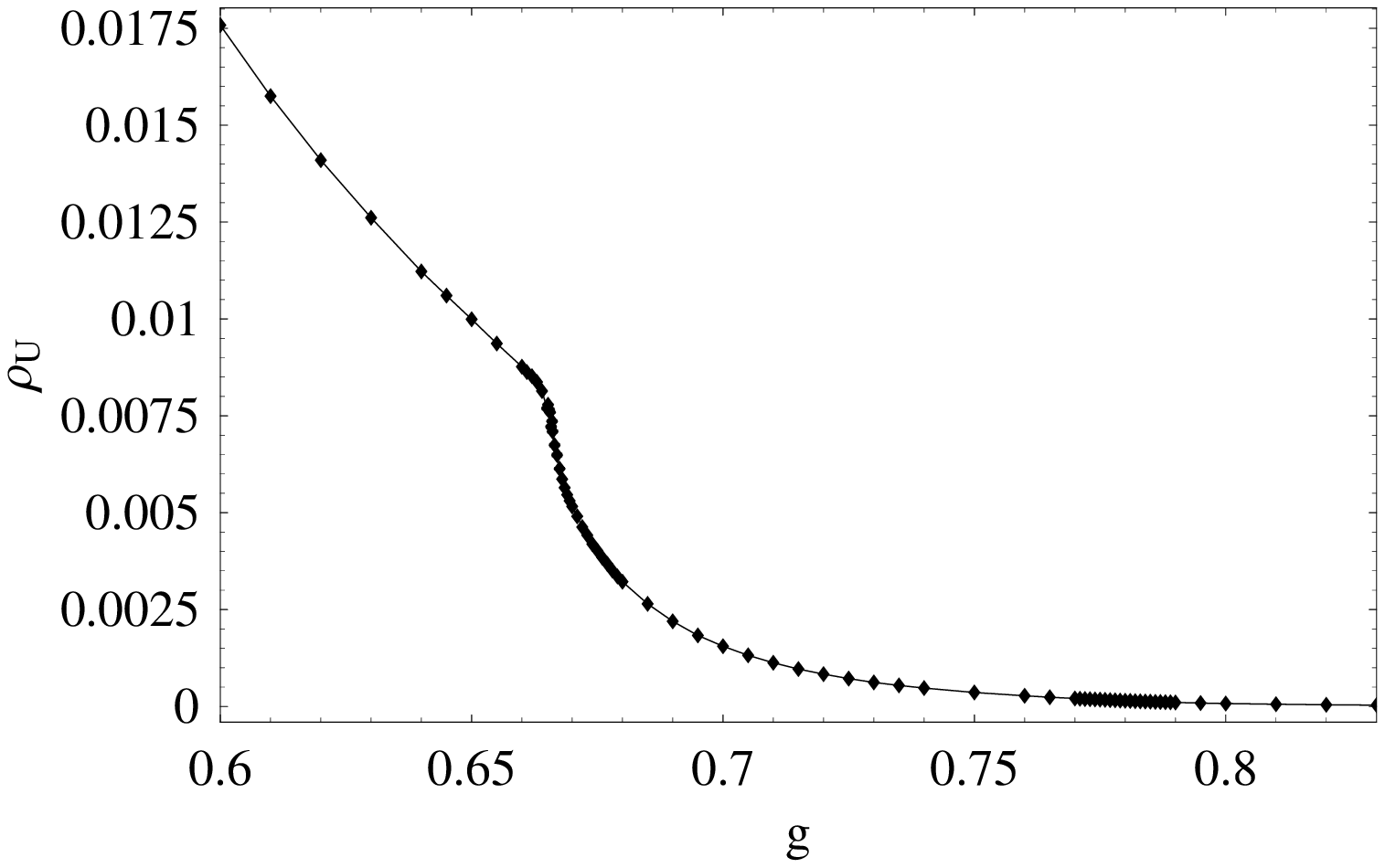}
\epsfxsize=7cm
\epsffile{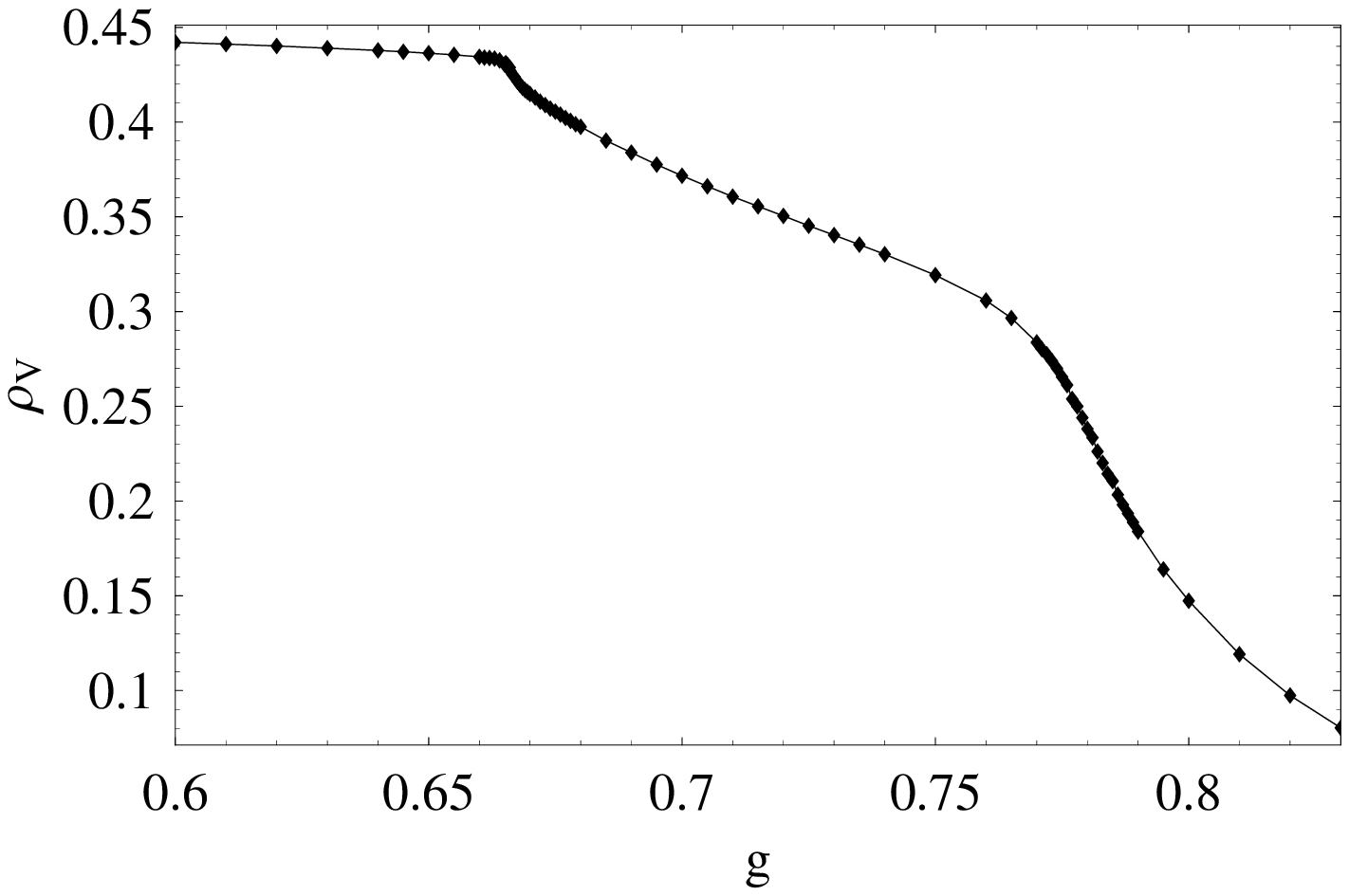}
\caption{U and V-monopole densities, $\rho_U$ and $\rho_V$, in the $(3+1)$D 
U-V model with $\alpha=\beta=5$.
}
\label{fig-inst}
\end{center}
\end{figure}

In Fig.\ref{fig-inst}, we show the calculation of the U and V-monopole 
densities.
We employ the definition of the monopole density in the lattice
gauge theory given in Ref.\cite{monopole}. 
Please remark that the $V$-field is a 3D ``vector field" and its monopole
configuration can be defined straightforwardly.
On the other hand, the $U$-field is a $(3+1)$ D ``vector field", and 
we define the U-monopole just focusing on its 3D spatial component. 

\begin{figure}[htbp]
\begin{center}
\epsfxsize=7cm
\epsffile{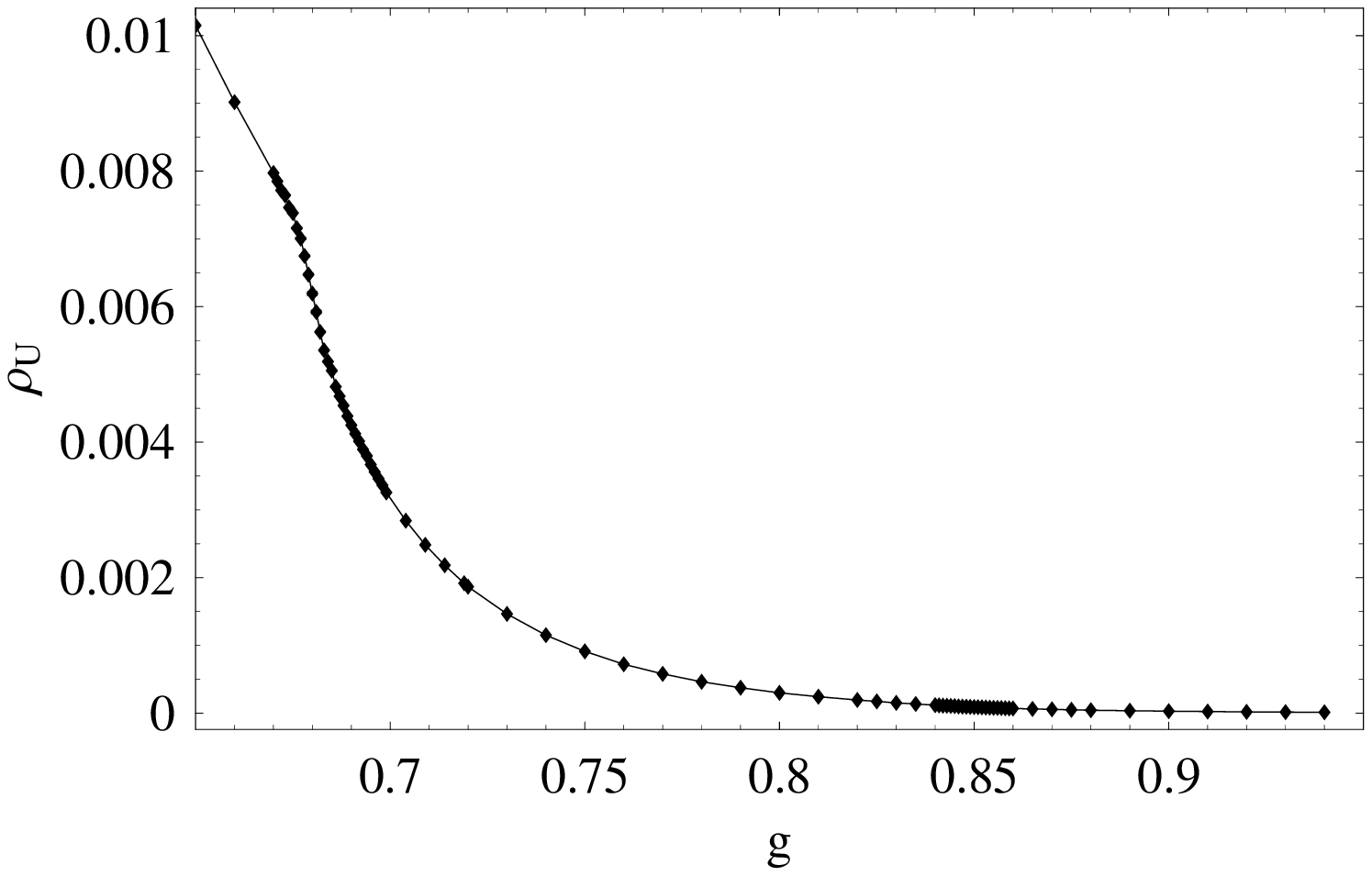}
\epsfxsize=7cm
\epsffile{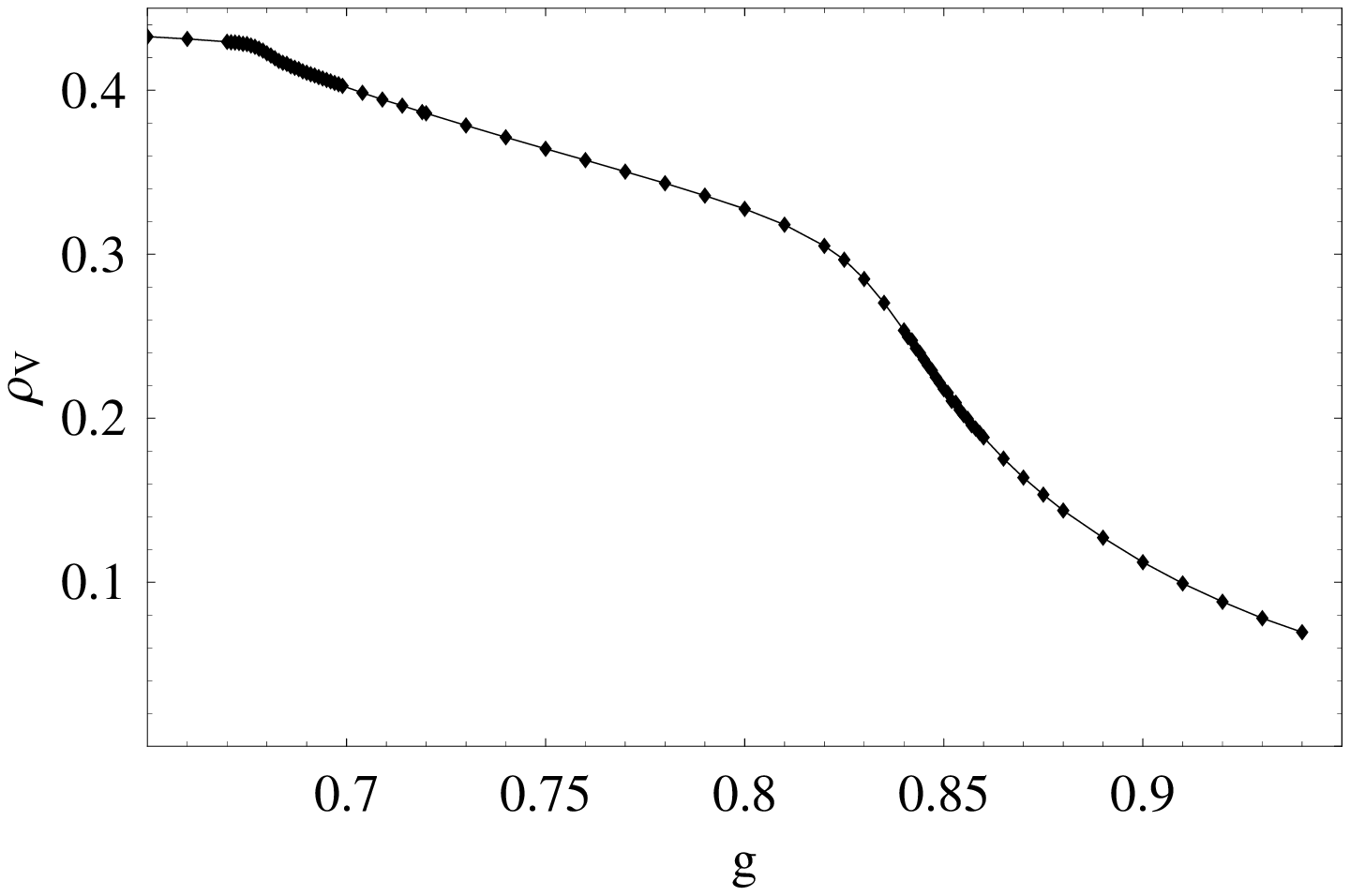}
\caption{U and V-monopole densities, $\rho_U$ and $\rho_V$, in the 
$(3+1)$D U-V model with $\alpha=\beta=10$.
}
\label{fig-inst2}
\end{center}
\end{figure}

From Fig.\ref{fig-inst}, it is obvious that the U-monopole density $\rho_U$ 
is very low in the whole parameter region of $g$ as the $U$-field is the 
{\em noncompact} gauge field.
It starts to decrease very rapidly at around the first phase transition point
$g_c\sim 0.67$.
At the second phase transition point $g'_c\sim 0.78$, $\rho_U$ is substantially
vanishing.
On the other hand, the V-monopole density $\rho_V$ starts to decrease at
$g_c\sim 0.67$, but it has a finite value between the first and second phase
transitions. 
After the second transition at $g'_c\sim 0.78$, $\rho_V$ decreases very 
rapidly.
Similar behavior is observed also in the case of $\alpha=\beta=10$ and the 
London limit.
(See Fig.\ref{fig-inst2}.)
This result indicates that a finite density of {\em short V-vortices 
connecting monopole and anti-monopole} survive as low-energy excitations 
in the parameter region between the first and second phase transition points.
From the above consideration, it is expected that the second phase transition
is a very specific one to the U-V model and it does not exist in the ordinary 
$s$-wave SC.

%---------------------------------------------------
%----------------------------------------------------

%---------------------------------------------------
%----------------------------------------------------

%%%%%%%%%%%%%%%%%%%%%%%%%%%%%%%%%%%%%%%%%%%%%%%%%%%%%%%%%%%%%%%%%%
\setcounter{equation}{0}
\section{Discussion}

In this paper, we have derived the effective gauge model of the Cooper-pair 
field sitting on links from the microscopic quantum  Hamiltonian by
the path-integral methods.
This lattice GL theory for unconventional SC is a lattice gauge model
with dual gauge fields. 
There the electromagnetic field $U_{x,\mu}$ is a {\em noncompact} U(1) gauge 
field, whereas the link Higgs field $V_{x,i}$ is a {\em compact} U(1)
gauge field.
Then we studied phase structure of the model mostly focusing on the QPT
at $T=0$ by means of the MC simulations.
For the $(2+1)$D case, we found that there exists a second-order phase 
transition for the normal to SC phases as the coupling constant $g$ 
controlling quantum fluctuations is increased.
The gauge-invariant gauge-boson mass start to develop at the critical 
coupling obtained by the measurement of the specific heat $C$.
On the other hand for the $(3+1)$D case, we found two second-order phase 
transitions.
From the results of the measurement of $M_G$ and the monopole densities 
$\rho_{U(V)}$, we identify the first one as the normal-SC phase transition and 
the second one as the transition of V-monopole suppression.

It is quite instructive to compare the above results to those obtained
in our previous papers, in which some related models are studied.
In Ref.\cite{UV1}, we investigated the {\em compact} U(1) U-V model 
in 3 and $(3+1)$D
in which $U_{x,\mu}$ is the compact U(1) gauge field.
In particular in section 6 of Ref.\cite{UV1}, we studied the QPT of 
the model and found 
that {\em a first-order} phase transition from the confinement to Higgs (SC) 
phases occurs.
On the other hand in the present paper, we found {\em two second-order} phase 
transitions for the {\em noncompact} U-V model in similar parameter regions
with the above.
This result can be understood as follows; in the compact U-V model, the $U$
and $V$-fields are (almost) symmetric though there is no time component of
$V$-field.
The monopole proliferation transition of the $U$ and $V$-gauge fields takes 
place simultaneously as actually observed in Ref.\cite{UV1}.
On the other hand in the noncompact case, there is the asymmetry between 
the $U$ and $V$ gauge fields. 
Therefore roughly speaking, the two second-order phase transitions in the
noncompact case coincide in the compact case and as a result a first-order
phase transition appears.
Phenomena similar to the above are expected to occur rather generally.
In fact in the very recent paper\cite{MHM}, we studied multi-flavor
Higgs U(1) lattice gauge models in various parameter regions
and found that when two second-order phase transitions coincide at certain
points in the parameter region of the model, a single first-order transition 
appears.
One may ask if there exist parameters controlling the distance between the two
second-order transitions in the present model.
We expect that the coefficient 
$c_3$ in the action (\ref{action}) controls the distance, e.g., the two peaks
get closer for larger value of $c_3$, as the topologically nontrivial 
excitations of the $V$-field are suppressed for large $c_3$.
This problem is under study and the result will be reported in a 
future publication.

Preliminary result of the numerical study on this problem seems to 
support the above expectation.
Furthermore, we found that the relative magnitude of two peaks in $C$ changes 
as the value of $c_3$ is varied.
For larger $c_3$, the second peak is getting larger.
If the two second-order transitions merge and as a result 
a single first-order phase transition appears as $c_3$ is getting large, 
interesting behavior of the {\em amplitude} of $V$-field 
$v\equiv |V_{x,j}|$ is expected to be observed.
From the behavior of $\rho_V$ in Figs.\ref{fig-inst} and \ref{fig-inst2}, 
it is expected that for $g<g'_c$ the potential term in $S_{\rm GL}$,
$-\beta v^2 +\alpha v^4$, determines the expectation value of $v$
as $\langle v \rangle \sim \sqrt{{\beta \over 2\alpha}}$ because the phase 
degrees of freedom of $V_{x,j}$ fluctuate strongly.
As $g$ is getting larger than $g'_c$, the other terms in $S_{\rm GL}$
start to contribute to $\langle v \rangle$ and as a result $\langle v \rangle$
is a sharp increasing function of $g$ near $g \sim g'_c$.
At the first-order phase transition pint, on the other hand, 
$\langle v\rangle$ is expected to exhibit a sharp discontinuity and have a
hysteresis loop as $\rho_V$ does. 
In any case, the above interpretation of the structure of the phase transitions
in the U-V models is useful and also interesting.

In Ref.\cite{UV2}, we studied {\em thermal phase transition} of the noncompact
U-V model in 3D.
There not only the electromagnetic gauge field $U_{x,i}$ but also 
the Cooper-pair field $V_{x,i}$ are put on all links of the cubic
lattice, i.e., the direction index $i=1,2$ and $3$.
We found {\em first-order} phase transitions from the normal to SC phases
in the parameter regions $c_2>0, \; d_2=0$ and $c_2=0,\; d_2<0$.
As the results obtained in the present paper for $c_2>0,\; d_2<0$
show the existence of the second-order QPT in 3D,
it is possible that the thermal phase transition becomes of second order
in the parameter region like $c_2>0,\; d_2<0$, i.e., there exist
tricritical points.
This problem is under study and results will be reported in a future 
publication.

Let us turn to the discussion on the QPT of the high-$T_c$ cuprates given
in the previous work.
In Ref.\cite{XY}, the QPT of the cuprates was studied by assuming that 
the critical
behavior can be described by a 3D quantum XY model whose Hamiltonian 
is given by,
\begin{equation}
H_{\rm XY}={1 \over 2}\sum_{x,i}\hat{n}_x{\cal V}_{x,y}\hat{n}_y
-{1 \over 2}\sum_{x,i}J_0\cos ({\varphi}_{x+i}-\varphi_x),
\label{HXY}
\end{equation}
where $\hat{n}_x$ is the number operator conjugate to the phase of the
Cooper pair $\varphi_x$ {\em on the site} $x$ of the cubic lattice, 
$[\hat{n}_x, \varphi_y]=i\delta_{x,y}$.
Detailed investigation on the model (\ref{HXY}) was given for the 
{\em short-range interaction} case, ${\cal V}_{x,y}=V_0\delta_{x,y}$,
and an effective GL theory in terms of the order parameter 
$\psi_x \sim e^{i\varphi_x}$ was derived.
Near the quantum critical point, the superfluid density $\rho_s$ was 
obtained as follows by means of a mean-field approximation neglecting
fluctuations of $\psi_x$,
\begin{equation}
\rho_s={8 \over 7}\Big({V_0 \over J_0}\Big)^2(J_0-V_0/4d).
\label{rhos}
\end{equation}
The critical point is identified as 
\begin{equation}
\Big({V_{0} \over J_0}\Big)_c=4d, 
\end{equation}
and for ${V_0 \over J_0}<4d \; ({V_0 \over J_0}>4d)$, the system is in the
SC (normal or insulating) phase.
Though we studied the case of the long-range interactions Eq.(\ref{V}) in 
this paper, we can qualitatively expect the relation like $V_0 \propto 1/g, 
\; J_0 \propto g$.
Then the above result concerning to the SC QPT is consistent with that 
obtained in this paper.
Furthermore as $M_G \propto \sqrt{1/\rho_s}$ in the SC phase,
Eq.(\ref{rhos}) predicts $M_G \propto \sqrt{(V_0/J_0)_c-(V_0/J_0)}
\propto \sqrt{g-g_c}$, where $g_c$ is the critical coupling constant.
The numerical results obtained in this paper qualitatively coincide
with this behavior of $M_G$ and are consistent with the experiments\cite{exp},
though in our model the Cooper pair is put 
on lattice links instead of sites and behaves like a compact gauge field.

%%%%%%%%%%%%%%%%%%%%%%%%%%%%%%%%%%%%%%%%%%%%%%%%%%%%%%%%%%%%%%%%%%
\newpage
\setcounter{equation}{0}
\appendix
\section{Derivation of the GL theory}
\renewcommand{\theequation}{A.\arabic{equation}} 

In this appendix, we shall prove Eqs.(\ref{SAtext}) and (\ref{HPsitext}).
We first prove Eq.(\ref{SAtext}).
To this end, let us consider the following quantity,
\begin{eqnarray}
H_A&=&\int dk \; d^2[e^{-2}\tilde{V}(k)-1]^{-1}
\tilde{A}_0(-k)\tilde{A}_0(k) 
+{1\over e^2}\sum_{x,i,j}(\dot{\phi}_{x,i}-eA_{x,0})
(\dot{\phi}_{x,j}-eA_{x,0})  \nonumber \\
&=&{1\over 2\pi}\int dk\sum_{i,j}\Bigg[\Big\{[e^{-2}\tilde{V}(k)
-1]^{-1}+1\Big\}\tilde{A}_0(-k)\tilde{A}_0(k) 
-{2\over e}\dot{\phi}_i(-k)\tilde{A}_0(k)+{1\over e^2}
\dot{\phi}_i(-k)\dot{\phi}_j(k)\Bigg] \nonumber \\
&=&{1\over 2\pi}\int dk\sum_{i,j}\Bigg[\Gamma(k)\Big(\tilde{A}_0(-k)-
{1\over e\Gamma(-k)}\dot{\phi}_i(-k)\Big)  
\Big(\tilde{A}_0(k)-{1\over e\Gamma(k)}\dot{\phi}_j(k)\Big)\nonumber \\
&&+\tilde{V}^{-1}(k)\dot{\phi}_i(-k)\dot{\phi}_j(k)\Bigg],
\label{HA}
\end{eqnarray}
where $\Gamma(k)=[1-e^{2}\tilde{V}^{-1}(k)]^{-1}$.
Therefore
\begin{equation}
e^{-S_{\phi}}=\int [dA_0]e^{-\int d\tau H_A}.
\label{SA}
\end{equation}

Next we shall prove Eq.(\ref{HPsitext}).
To this end, let us consider the following integral,
\begin{eqnarray}
H_V&=&{1\over 2}\sum_{x,i\neq j}\Big[J^{-1}e^{iA}V^\ast_{x,i}
V_{x+j,i}-V_{x,i}e^{-i\phi_{x,i}}+\mbox{c.c.}\Big] \nonumber \\
&=&{1\over 2}\sum_{x,i\neq j}\Big[J^{-1}e^{iA} 
 \Big(V^\ast_{x,i}-Je^{-iA}
e^{-i\phi_{x+j,i}}\Big)\Big(V_{x+j,i}-Je^{-iA}
e^{i\phi_{x,i}}\Big) \nonumber \\
&&\;\;\;\hspace{1cm} -Je^{i(\phi_{x,i}-\phi_{x+j,i}-A)}+\mbox{c.c.}\Big],
\label{Psi2}
\end{eqnarray}
where $A\equiv A_{x+i,j}+A_{x,j}$.
Therefore
\begin{equation}
e^{-\int d\tau H_t}=\int [dV^\ast dV]e^{-\int d\tau H_V}.
\label{HPsi}
\end{equation}

%%%%%%%%%%%%%%%%%%%%%%%%%%%%%%%%%%%%%%%%%%%%%%%%%%%%%%%%%%%%%%%%%%

%%%%%%%%%%%%%%%%%%%%%%%%%%%%%%%%%%%%%%%%%%%%%%%%%%%%%%%%%%%%%%%%%%%

\end{document}